\PassOptionsToPackage{table}{xcolor}
\documentclass[sigconf]{acmart}
\usepackage{tikz}
\usetikzlibrary{positioning,calc,arrows.meta}
\usepackage{multirow}
\usepackage{array}
\usepackage{booktabs,tabularx}
\usepackage{graphicx}
\usepackage{subcaption,acmart-taps}
\usepackage{makecell}
\newcolumntype{Y}{>{\raggedright\arraybackslash}X}
\usepackage{newunicodechar}
\newunicodechar{（}{(}
\newunicodechar{）}{)}
\usepackage{pgfplots,colortbl,xcolor}
\pgfplotsset{compat=1.18} 
\usepackage{pdflscape}
\newcolumntype{P}[1]{>{\raggedright\arraybackslash}p{#1}}
\renewcommand{\arraystretch}{1.15} 

\copyrightyear{2026}
\acmYear{2026}
\setcopyright{cc}
\setcctype{by}
\acmConference[CHI '26]{Proceedings of the 2026 CHI Conference on Human Factors in Computing Systems}{April 13--17, 2026}{Barcelona, Spain}
\acmBooktitle{Proceedings of the 2026 CHI Conference on Human Factors in Computing Systems (CHI '26), April 13--17, 2026, Barcelona, Spain}
\acmPrice{}
\acmDOI{10.1145/3772318.3790457}
\acmISBN{979-8-4007-2278-3/2026/04}

\AtBeginDocument{%
  }
    
\begin{document}

\title{“It Depends”: Re-Authoring Play Through Clinical Reasoning in Wearable AR Rehab Games }
\author{Binyan Xu}
\authornote{These authors contributed equally to this work.}
\affiliation{%
  \institution{Bouvé College of Health Sciences, Northeastern University}
  \city{Boston}
  \state{Massachusetts}
  \country{United States}
}
\email{xu.biny@northeastern.edu}
\author{Wei Wu}
\authornotemark[1]
\affiliation{%
  \institution{Ghost Lab, Northeastern University}
  \city{Boston}
  \state{Massachusetts}
  \country{United States}
}
\email{wu.w4@northeastern.edu}
\author{Soonhyeon Kweon}
\affiliation{%
  \institution{Bouvé College of Health Sciences, Northeastern University}
  \city{Boston}
  \state{Massachusetts}
  \country{United States}
}
\email{kweon.s@northeastern.edu}

\author{Casper Harteveld}
\affiliation{%
  \institution{College of Arts, Media and Design, Northeastern University}
  \city{Boston}
  \state{Massachusetts}
  \country{United States}
}
\email{c.harteveld@northeastern.edu}

\author{Leanne Chukoskie}
\affiliation{%
  \department{Bouvé College of Health Sciences; College of Arts, Media and Design}
  \institution{Northeastern University}
  \city{Boston}
  \state{Massachusetts}
  \country{United States}
}
\email{l.chukoskie@northeastern.edu}

\begin{abstract}
Augmented reality (AR) games hold promise for rehabilitation, yet most remain confined to laboratory studies with limited clinical uptake. Recent advances in spatial computing, especially lightweight, glasses-form-factor AR, create a timely opportunity to embed rehabilitative play into clinical practice and daily contexts. To investigate this potential, we systematically reviewed 132 applications and conducted playtesting with 14 licensed physical therapists. Our analysis revealed three ways therapists re-authored AR games: co-authored play (reshaping movements, progressions, and difficulty), situated play (adapting across specialties, conditions, and contexts), and dual play (mediating both physical recovery and psychological support). We reframe therapists’ frequent phrase—“It depends”—as a generative design principle. This study contributes a clinical reasoning–based framework and design principles and guidelines for creating personalized, situated forms of play that align with therapists’ everyday workflows and inform future lab-to-clinic translation.
\end{abstract}



\begin{teaserfigure}
  \centering
  \includegraphics[width=15cm, height=7cm]{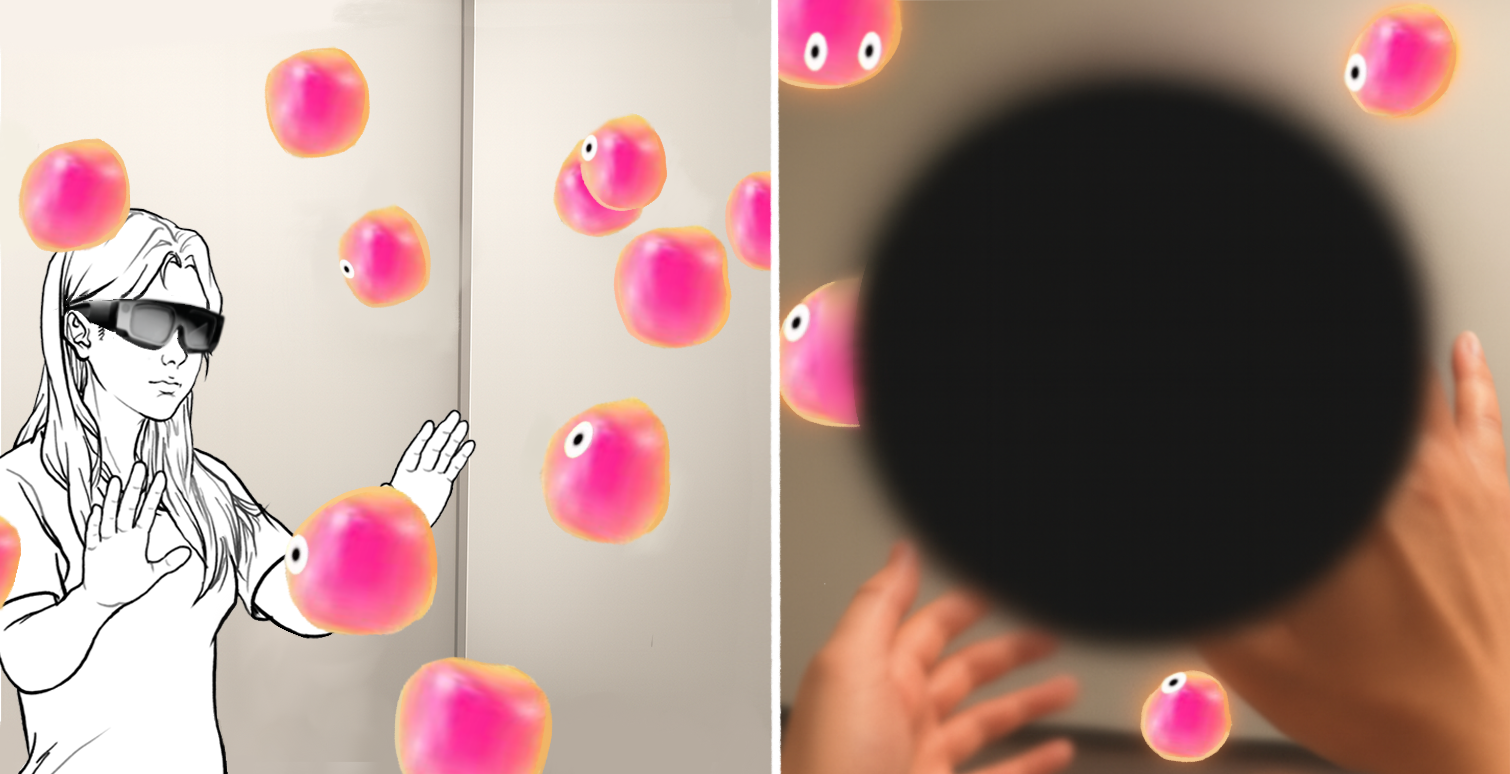}
  \caption{Teaser illustration of a stroke survivor with central vision loss engaging in AR-based rehabilitation. 
  The left panel (third-person view) shows a patient wearing AR glasses, surrounded by playful floating game objects representing therapeutic tasks. 
  The right panel (first-person perspective) simulates the patient’s visual experience: a central black circular scotoma forces the user to rely on head scanning to detect and interact with peripheral objects, enabling everyday activities.}
  \Description{This teaser figure depicts the imagined rehabilitation scenario described by PT05. 
  On the left, a third-person view of a stroke survivor wearing AR glasses interacting with floating game objects. 
  On the right, a first-person simulation of central vision loss: a black circle blocks the center, requiring head movements to scan the periphery and complete tasks.}
  \label{fig:teaser}
\end{teaserfigure}

\maketitle

\section{Introduction}

A stroke survivor sits in the clinic, their central vision lost (see Figure~\ref{fig:teaser}). To see the room, they must turn their head, scanning each fragment with effort. Their therapist hangs colored ribbons as visual targets, but frustration quickly sets in. Training feels punitive, and motivation fades. Months later, during our study, this same physical therapist (PT) tried a spatial elimination game using lightweight AR glasses. Almost before she removed the glasses, she asked: “Can I bring this to my clinic tomorrow?” To her, the familiar, low-burden form factor of glasses, combined with a playful spatial game unfolding directly in the real environment, offered more than training.It suggested a clinic-ready way to transform effortful repetition into cooperative engagement.

This case illustrates how physical therapists (PTs) routinely reframe therapy as \emph{play} in clinical settings. Sometimes this takes the form of inflating a glove into a volleyball in an ICU ward; at other times, it involves subtle adjustments of posture or attention to make routine movement engaging. Increasingly, it includes experimentation with digital tools. Across these practices lies a shared aim: reshaping difficult or repetitive work into something patients can meaningfully engage with. While a growing body of lab-based research suggests AR can improve engagement and outcomes in rehabilitation ~\cite{Gil2021AR,alissa2012ar}, PTs often respond with a familiar phrase when asked how such systems would fit into real clinical care: \emph{“It depends.”} This response is not evasive. It reflects the flexible, case-by-case logic at the heart of clinical reasoning. At the same time, it raises a central design question: \textbf{How, under what circumstances, and with what design principles can lightweight wearable AR games be meaningfully integrated into everyday physical therapy practice?}

Physical-therapy games are widely recognized for supporting long-term patient adherence~\cite{Doumas2021Serious,CortesPerez2021LeapMotion,Pope2017ActiveVideoGames,UnibasoMarkaida2022CommercialGames}. In HCI, much of the literature pairs co-design with PTs with small-scale usability studies or short-term deployments~\cite{kaminer2014assets,Hymes2021CHIPlay,Bu2022HAI,duval2022taac,bayrak2021ozchi}. Far fewer studies explain how such games can be sustained in safety-critical, equipment-rich clinical sessions, where therapists must continuously supervise, guard, cue, and adapt tasks to the individual~\cite{Jung2020RehabGames,Richards2005OTStroke}. This leaves a core gap: reusable, PT-centered design principles that fit routine workflows while supporting patient-specific adaptation.

Prior studies of immersive and headset-based systems highlight persistent clinical workflow constraints, including the need for supervision, communication, and full environmental awareness during therapy with real equipment~\cite{chiu2024novelty,Gil2021AR,Hsieh2018VRAR,Hardeman2024RemoteARProtocol}. We build on these findings by exploring lower-burden, see-through wearable AR form factors that address complementary design needs in routine clinical practice. Throughout this paper, we focus on lightweight, glasses-form AR devices (e.g., Snap Spectacles) and refer to them as \emph{AR glasses}. By embedding game elements into ongoing therapy tasks, AR glasses can support in-situ cueing and functional practice for “in-the-flow” use during standard sessions~\citep{Liu2025LowerLimbAR}. Early studies report positive acceptance and motivation in some contexts~\cite{son2022development,Balloufaud2025ARExergames}. Yet key design challenges for AR rehabilitation games remain underspecified. Prior VR and AR work has connected augmentation, embodied action, and task embedding~\cite{Herrera2025VR,Chen2019VAMR,Sun2025EmbodiedGrasping}, but typically at the level of individual applications rather than a reusable, PT-centered design framework.

We argue that effectively integrating AR glasses into rehabilitation requires coordinating these layers through \emph{clinical reasoning}, the adaptable framework PTs use to make decisions under uncertainty and complexity~\cite{Edwards2004,Huhn2019}. Clinical reasoning shapes how therapists tailor rehabilitation plans to individual patients and position interventions within routine clinical workflows, allowing game-based systems to amplify engagement while preserving safety and continuity of care. We propose clinical reasoning not only as a way to understand PTs’ practice, but as a \emph{design lens for AR rehabilitation games}.

To explore this, we analyzed the full set of 132 applications in the developer library of Snap Spectacles, the only publicly released wearable AR glasses platform at the time of our study. From this corpus, we identified 21 applications with clear movement relevance and selected four for in-depth study. We then conducted playtesting with think\textendash aloud protocols and follow\textendash up interviews with ten licensed PTs (1--37 years' experience). Our analysis surfaced three propositions: \textbf{play is co\textendash authored}, as therapists actively shape its form and progression; \textbf{play is situated}, as its meaning shifts across bodies, specialties, and clinical contexts; and \textbf{play is dual}, as it mediates both physical and psychological dimensions of rehabilitation.

This paper contributes: 
(1) a clinical reasoning–based framework explaining how licensed PTs reinterpret and reshape AR rehabilitation games using lightweight AR glasses; 
(2) three propositions (co-authored, situated, and dual) that articulate how PTs adapt game mechanics to different bodies, contexts, and treatment goals; and 
(3) design principles that support the creation of adaptable rehabilitation games aligned with PTs’ everyday practice and professional decision-making.

\section{Related Work and Background}
\subsection{XR Rehabilitation and Clinical Translation Challenges}
Extended Reality (XR) technologies have shown promise in rehabilitation, especially for neuro- and
motor-recovery contexts~\citep{Akbar2024,Hariharan2024,BulleSmid2023XRABI,Gaballa2022}.
Meta-analyses and Cochrane reviews report that, for selected populations and outcomes, XR combined with conventional therapies can improve upper-limb function, balance, and performance of daily activities~\citep{Laver2017,Phan2022,Demeco2023IVR,Khan2023VRStroke}.
Scoping reviews further highlight XR’s potential flexibility to support task-specific training and multisensory feedback that may enhance engagement and adherence~\citep{matamala2019immersive,Liu2025LowerLimbAR}.
Taken together, this body of work suggests that XR can serve as a useful complement to traditional rehabilitation under controlled or well-supported conditions, while also underscoring that effects are heterogeneous across patient groups, contexts, and implementation models. At the same time, multiple reports note that XR adoption in everyday practice remains uneven.
Many interventions are evaluated as lab-based prototypes or short-term clinical pilots, and there are relatively few accounts of multi-year, routine integration into therapy workflows~\citep{Glegg2018,jones2023xrprimarycare}.
Documented barriers include hardware cost and complexity, infection-control requirements, and limited therapist training, among others~\citep{Khan2024Immersive,SimonLiedtke2022UniversalXR,Kourtesis2024MultimodalXR,jones2023xrprimarycare}.
Implementation-focused studies emphasize that translation from lab to clinic is a socio-technical challenge shaped by workflow, organizational readiness, reimbursement structures, and patient heterogeneity, rather than a purely technical problem~\citep{Greenhalgh2017,Naqvi2024}.

These translation challenges resonate with issues described in the broader field of \emph{digital therapeutics (DTx)}.
DTx systems (software-based medical interventions formally validated for regulatory approval) typically emphasize protocol-driven delivery and regulatory fidelity.
While this can yield robust, standardized interventions, prior work has noted that such systems may offer limited support for the kinds of in-the-moment adjustments that characterize physical therapy practice, where therapists continually tune difficulty, context, and engagement strategies in situ~\citep{Wang2023DigitalTx}.
We draw on these critiques to motivate a complementary perspective in which therapists are treated as co-authors of play and AR rehabilitation games are designed to enable adaptable, situated tailoring.
By situating AR games within therapists’ clinical reasoning, we propose a set of design levers for flexibility, co-authorship, and context-sensitive play that can inform future efforts to narrow the lab-to-clinic gap. 

\subsection{AR Glasses and Situated Interaction}
AR glasses afford \emph{situated} interaction by embedding digital content into everyday contexts.
Foundational surveys established AR as an overlay medium~\citep{Azuma1997,Billinghurst2015}, and recent progress in head-worn systems---including lightweight smartglasses and optical see-through headsets (e.g., HoloLens, Magic Leap, Xreal)---supports hands-free, in-situ use in clinical and home settings~\citep{Rauschnabel2019,Doughty2022Augmenting,Tran2023WearableAR,Palumbo2022HoloLens,Gsaxner2023HoloLens,Gorman2022ARStroke,patil2022integration,Gil2021AR,pereira2020handrehab,Xu2025PTMovementLogicsSpectacles}.
Together, these advances have expanded rehabilitation use cases beyond the laboratory.

Clinical studies show that AR glasses can situate therapy within real-world tasks (e.g., gait cueing, dual-task training, balance assessment)~\citep{Hardeman2024,rosenfeldt2025augmented}. At the same time, embedding AR in real contexts raises design obligations around hygiene and infection control in hospital wards~\citep{Hsieh2018VRAR,Kwon2022SmartHospital}, patient safety during movement, and cognitive load.
Compared to fully immersive VR headsets, which typically construct closed virtual tasks behind an occluding display, AR glasses keep patients situated in their actual environment and social field.
Accordingly, in this work, we treat AR rehabilitation as less about authoring self-contained virtual tasks and more about coordinating three coupled layers---\emph{visual augmentation}, \emph{embodied action}, and \emph{environmental embedding}---as they unfold in practice.

\subsection{Games and Play in Rehabilitation}
Games have long been explored as motivational tools in rehabilitation, but their role extends beyond engagement. 
Meta-analyses report that video games can improve adherence, enjoyment, and practice intensity compared to conventional therapy \citep{Doumas2021Serious,CortesPerez2021LeapMotion,Pope2017ActiveVideoGames,UnibasoMarkaida2022CommercialGames,Anderson2013YouMove,wu2025ghostgait}. 
However, HCI studies increasingly emphasize that games in rehabilitation are not fixed artifacts but practices that are re-authored with clinicians and patients.

To clarify our terminology, we position \emph{embodied interaction mechanics} in relation to existing game-design and HCI constructs.
In game design, \emph{core mechanics} describe the primary repeated actions and rules that structure moment-to-moment play \citep{Fullerton2018GameDesignWorkshop}. 
Building on this, HCI introduces \emph{embodied core mechanics} to emphasize that, in movement-based play, these repeated actions are inherently bodily and are co-shaped by rules, artifacts, and socio-spatial arrangements \citep{MarquezSegura2016EmbodiedCoreMechanics,MarquezSegura2016Bodystorming}. 
In this paper, we use \emph{embodied interaction mechanics} to denote the embodied core actions and feedback loops that a rehabilitation game affords, and that can be re-parameterized through clinical reasoning (e.g., movement quality, range, pacing, and functional context), aligning with recent embodied design work in physiotherapy and rehabilitation contexts \citep{VegaCebrian2024Movits}.

Research within CHI and CHI PLAY shows that therapists do not merely “use” games—they actively adapt and appropriate them to meet clinical goals. Drawing on Self-Determination Theory,~\citet{aufheimer2023motivation} explain how specific game features can support intrinsic motivation in physical therapy. Moving from features to design process, \citet{Cheng2017Leveraging} identifies patterns that scaffold designer–therapist collaboration in movement-based games. Likewise,~\citet{Hymes2021} shows that aphasia rehabilitation games often require reworking rules and feedback to honor clinical priorities and patient identities. Extending these perspectives,~\citet{chiu2024novelty} traces how VR exergames move from novelty to clinical practice when therapists themselves mediate this translation, further illustrating that play in rehabilitation is co-authored—negotiated among clinicians, designers, and patients—rather than pre-defined by software.

Broader HCI research on exertion games and embodied interaction provides conceptual grounding for this perspective. 
Exertion game frameworks emphasize how effort, rules, and social context shape bodily play~\citep{Mueller2011,Mueller2003}, while design theory situates play as appropriation, adaptation, and negotiation rather than a fixed set of features. 
Motivational theories such as Self-Determination Theory remain influential~\citep{RyanDeci2000,Przybylski2010}, but most rehabilitation games operationalize them superficially—tying progression to abstract scores rather than to therapeutic parameters like movement quality or contextual function. 
What remains underdeveloped is a framework that connects play design to the adaptable logics therapists use in clinical reasoning.

\subsection{Clinical Reasoning in Physical Therapy}
\begin{figure}[!h]
    \centering
    \includegraphics[scale=0.2]{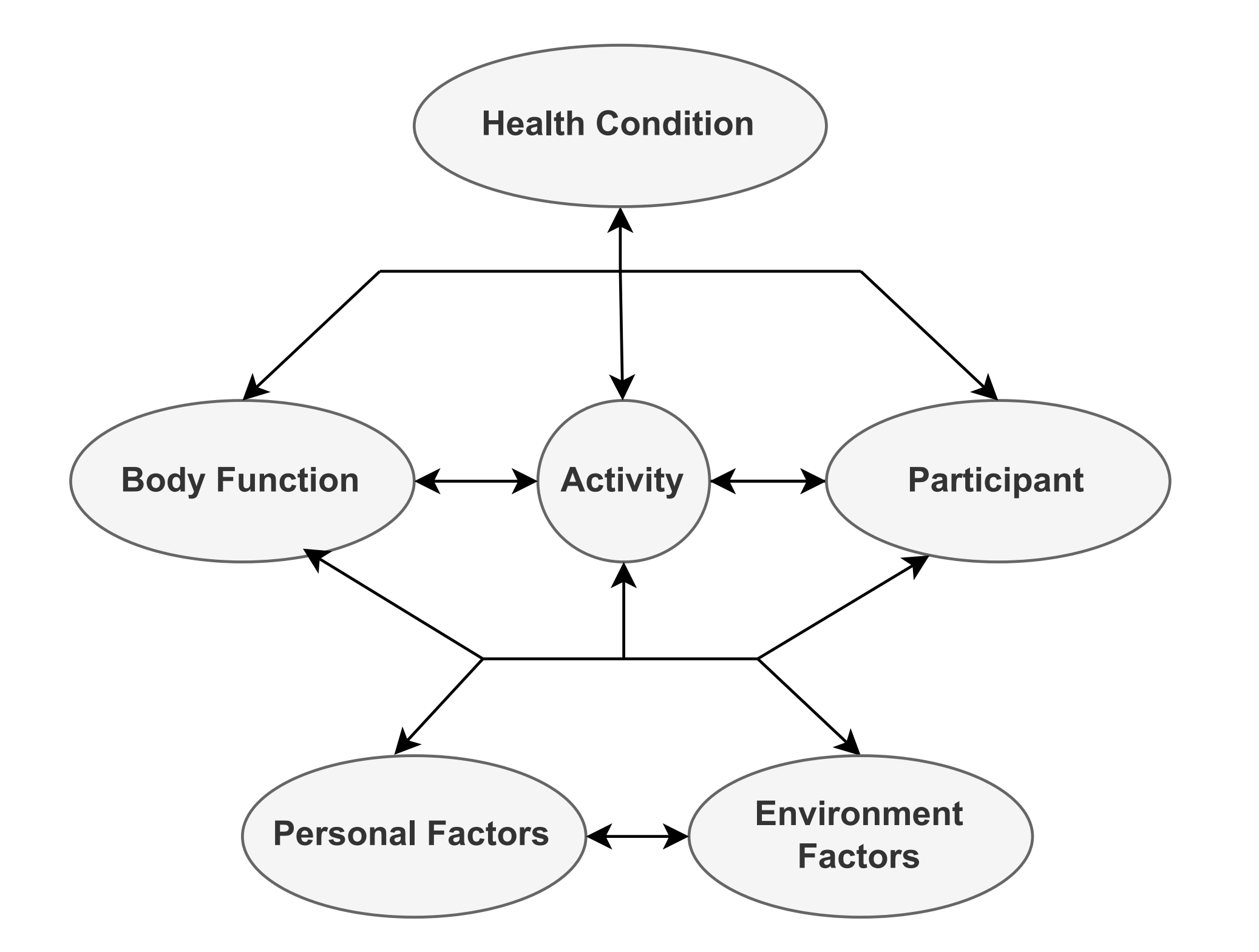}
    \caption{ICF model (WHO, 2001) linking Body Functions \& Structures, Activities, Participation, and Contextual Factors (Environmental \& Personal).}
    \Description{Diagram of the ICF model showing Health and Functioning at the center with surrounding components: Body Functions and Structures, Activities, Participation, and Environmental and Personal Factors, connected by arrows.}
    \label{fig:model}
\end{figure}
Clinical reasoning (CR) has been described as the process by which physical therapists individualize care amid patient and contextual variability~\citep{Edwards2004,Huhn2019,Musolino2024Clinical}. It provides a structured yet flexible way to act under uncertainty—the same adaptable logic needed when moving rehabilitation games from controlled laboratory studies into diverse clinical settings.

Early accounts of medical problem solving framed reasoning as hypothetico–deductive generation and testing of candidate explanations~\citep{Elstein1978,Elstein1990}, complemented by non-analytic strategies such as pattern recognition and illness scripts~\citep{Schmidt2015}. Dual-process theories integrate these modes, proposing that analytic and non-analytic reasoning operate in tandem and are coordinated through reflection and metacognition~\citep{Eva2005,Norman2010,Schon1987}.

Within PT, CR is inherently multidimensional and situated in daily practice. Ethnographic studies show how therapists weave diagnostic, narrative, interactive, pragmatic, and ethical considerations to align actions with patient goals and constraints~\cite{Huhn2019,Edwards2004,Jones2018Clinical}. The World Health Organization’s \emph{International Classification of Functioning, Disability and Health (ICF)} further structures this reasoning by linking body functions and structures with activities, participation, and contextual factors~\citep{WHO2001,Schenkman2006,Atkinson2011}. As a concrete instantiation of a biopsychosocial approach to care, the ICF shifts attention away from etiology alone toward what people are able to do, in which contexts, and with what supports or barriers. In this sense, it functions both as a holistic framework for thinking about care and as a communication scaffold that translates complex functional contexts into structured, portable summaries across teams and settings~\citep{WHO2001,Schenkman2006,Atkinson2011}.

In this work, we do not adopt the ICF as a full diagnostic taxonomy, but as a practical bridge between PTs and AR game developers and designers. Because the ICF is already embedded in everyday PT documentation and communication, it offers an immediately legible way for therapists to map digital game elements onto familiar categories (e.g., whether a mechanic primarily targets body functions, activity-level tasks, or contextual constraints). Rather than proposing an entirely new vocabulary, we therefore adapt the ICF as a shared scaffold that aligns therapist-tunable game parameters (e.g., movement range, dual-task demands, environmental setups) with clinical goals, foregrounding our translational focus.

\section{Positionality}
Our research team is interdisciplinary, working at the intersection of physical therapy, games, and HCI research. Two authors are based in a Department of Physical Therapy, two are in a program in games/game science with prior experience in rehabilitation-game research, and one author holds a joint appointment bridging both areas. The team also includes a licensed PT trained within the U.S. PT system, who contributed clinical expertise to protocol development, safety review, and interpretation of therapists’ reasoning. We were attentive to how our clinical and HCI backgrounds might shape what we noticed; to mitigate this, we used structured rubrics during game selection and analysis, open-ended interviews, and iterative multi-researcher coding to keep interpretations grounded in participants’ accounts. No researcher held supervisory or clinical authority over participants, and the AR games were presented as exploratory design probes rather than clinical interventions or prescriptions.

\section{Game Selection Study}
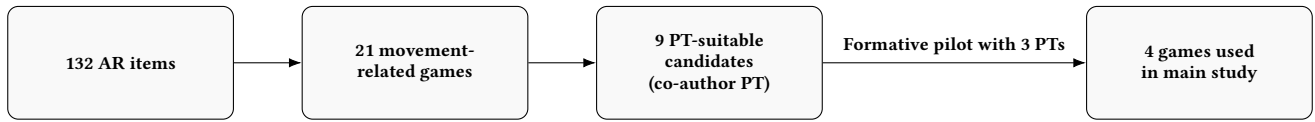
\begin{figure*}[t]
    \centering
    \begin{tikzpicture}[
        >=Latex,
        every node/.style={font=\footnotesize\bfseries}, 
        box/.style={
            rectangle,
            rounded corners,
            draw,
            align=center,
            fill=gray!5,
            minimum height=1.5cm,
            minimum width=3.0cm
        }
    ]
    \node[box]                          (n1) {132 AR items};
    \node[box, right=0.9cm of n1]       (n2) {21 movement-\\related games};
    \node[box, right=0.9cm of n2]       (n3) {9 PT-suitable\\candidates\\(co-author PT)};
    \node[box, right=3.5cm of n3]       (n5) {4 games used\\in main study};

    \draw[->] (n1) -- (n2);
    \draw[->] (n2) -- (n3);
    \draw[->] (n3) -- node[above, align=center] {\textbf{Formative pilot with 3 PTs}} (n5);
    \end{tikzpicture}%
    \caption{Game selection and refinement pipeline from the initial pool of 132 AR mini-games to the four games used in the main study.}
    \Description{Flow diagram illustrating the game selection and refinement process. Four rectangular nodes represent successive stages, connected by arrows from left to right: an initial set of 132 AR items, a filtered set of 21 movement-related games, a further refinement to 9 PT-suitable candidates co-authored with a physical therapist, and a final selection of 4 games used in the main study. An annotation above the final transition indicates a formative pilot involving three physical therapists.}

    \label{fig:game-selection-flow}
\end{figure*}

We engaged \textbf{14 licensed PTs} overall, including \textbf{4 PTs} in the \textbf{game-selection phase}: \textbf{1 PT} served as a corpus co-reviewer during instrument development (\emph{methods support only; not a study participant}), and a \textbf{formative pilot} with \textbf{3 PTs} was used to refine the game set and study materials (\emph{pilot data excluded from analysis}). The \textbf{main study} included \textbf{10 PTs} (\emph{analyzed}). The study protocols for the game selection and main study received Institutional Review Board (IRB) approval. This process is summarized in Figure~\ref{fig:game-selection-flow}.

\subsection{Catalog and screening.}
To ground our study in the current AR ecosystem, three researchers (an HCI researcher, an AR developer, and a licensed physical therapist) independently screened and coded Snap \emph{Spectacles} content as of April 1, 2025. The corpus comprised \textbf{104 commercial AR Lenses} and \textbf{28 open-source repositories} on GitHub (\textbf{132 items} total), which we classified into eight genres: Utility (39), Advertising (3), Feature Prototype (15), AI Assistant (6), Educational (14), Games (52), and Social (3). Disagreements were resolved through discussion. Within the Game genre, we excluded card and chess games, narrative titles, and other experiences that are not movement-based. Of the remaining games, \textbf{21} exhibited clear body-movement interaction mechanisms. A full breakdown of these 21 games—covering interaction style, spatial layout, and movement demands—is presented in Appendix~\ref{app:game_eval}, Table~\ref{tab:pt_eval_21games}.

The \textbf{21 movement-related games} were then evaluated by the licensed PT co-author using a structured rubric with four dimensions: (a) \emph{body-part involvement} (e.g., upper limb, lower limb, fine motor), (b) \emph{functional similarity} to standard PT assessments or exercises, (c) \emph{technical fidelity} (e.g., tracking stability, spatial anchoring, feedback clarity), and (d) \emph{clinical applicability} (e.g., safety considerations, seated/standing variants, supervision needs). The resulting categorization for all 21 games is summarized in the Appendix~\ref{app:game_eval}, Table~\ref{tab:pt_game_groups}.
Based on this PT-led review, we identified \textbf{nine games} with clear potential for PT clinical practice. These nine titles were organized into \textbf{four PT training categories}. Within each category, games are listed in descending order of clinical preference, with earlier entries treated as primary candidates and later ones retained as backups. The resulting grouping and within-category ordering are summarized in Table~\ref{tab:pt_game_groups_table}.

\begin{table}[t]
\footnotesize
\renewcommand{\arraystretch}{1.15}
\setlength{\tabcolsep}{5pt}
\centering
\caption{PT-informed grouping and within-group ordering of the selected Snap AR games.}
\label{tab:pt_game_groups_table}
\begin{tabularx}{\columnwidth}{
  >{\raggedright\arraybackslash}p{3cm}
  >{\raggedright\arraybackslash}X
}
\toprule
\textbf{PT-Relevant Category} & \textbf{Games (preferred $\rightarrow$ backup)} \\
\midrule
Upper Extremity (UE) & ActionBall; Beat Boxer; Whack-a-Mole \\
Whole Body & Squishy Run; Pop Game \\
Neuro & LEGO Bricktacular; 3 Dots--Pinch \\
Real-Tool Integration & Ball Games; Darts \\
\bottomrule
\end{tabularx}
\end{table}

\subsection{Formative Pilot}

With the 21 movement-related games, we conducted a formative pilot with \textbf{three research-focused licensed PTs} to (1) select and validate the \textbf{final four games} for the main study (one per PT-relevant category) and (2) refine our study protocol and interview guide. Participant demographics for the formative pilot are summarized in Table~\ref{tab:pilot-pt-demographics}.

\begin{table}[h]
\centering
\caption{Pilot study PT participant demographics (n=3).}
\label{tab:pilot-pt-demographics}
\begin{tabular}{c c c c c}
\toprule
\textbf{PT ID} & \textbf{Specialty} & \textbf{Years} \\
\midrule
PPT01 & Sports medicine; Geriatrics & 4--6 \\
PPT02 & Orthopedics & 4--6 \\
PPT03 & Neuro & 7--10 \\
\bottomrule
\end{tabular}
\end{table}

Each pilot participant completed a 60-120 minute playtesting session in a laboratory within the physical therapy department at Northeastern University. After completing a brief pre-study questionnaire on clinical specialty and background, participants played 9 Snap AR games across 4 PT-relevant categories. Games were played category by category; within each category, the order of games was randomized. All sessions were audio- and video-recorded. Participants were prompted to think aloud during gameplay, followed by a short semi-structured interview after each category. Interview topics covered (1) \emph{movement relevance}—whether in-game actions resembled therapeutic activities and how PTs might structure progression (e.g., “What movements seem therapeutically useful?”); (2) \emph{engagement \& cognition}—links between game elements, cognitive load, and patient motivation (e.g., “What cognitive functions do these games engage?”); and (3) \emph{clinical suitability}—appropriate patient groups, rehab stages, and potential adaptations for real-world use (e.g., “Who could benefit from this game?”). After completing all categories, we conducted a cross-game comparison to elicit relative judgments and identify missing therapeutic features (e.g., “Which game has the strongest rehab potential?”). The full pilot protocol is provided in the Appendix. Pilot participants were compensated at \$20/hour.

\subsection{Final Game Selection after Pilot}

Across the structured play sessions and debrief interviews, we converged on \textbf{four representative games} that collectively offered stable tracking, clear mechanics, and strong relevance to PT training goals. The final main-study set consisted of: \textit{3 Dots--Pinch} (fine motor + ambulatory positioning), \textit{Squishy Run} (walking + reach), \textit{Ball Games} (lower-limb coordination + physical object tracking), and \textit{ActionBall} (upper-limb ROM + whole-body coordination). These four games are described in detail in the main study section below. Three of these titles aligned with the PT co-author’s prior rubric-guided review. In the Neuro category, however, we selected \textit{3 Dots--Pinch} in place of \textit{LEGO Bricktacular}. Pilot feedback indicated that LEGO’s AI-driven voice/color-request mechanic (e.g., verbally specifying a desired block color to retrieve it) could introduce unnecessary cognitive confusion for patient populations, whereas \textit{3 Dots--Pinch} provided a cleaner motor-focused interaction. 

\textit{The pilot also informed two protocol refinements}. First, we adjusted our questioning about cognition: an initial open-ended prompt (e.g., “What cognitive functions are these games related to?”) occasionally led participants to freeze, likely because it resembled a pop quiz. We therefore reframed it into a more conversational clinical style (e.g., “Do you think this game relates to any cognitive functions?”), which elicited richer reflections and better matched how PTs speak with patients. Second, we introduced a seated option for post-game interviews. Although seating was not controlled in the pilot, we observed that participants generated more clinically grounded scenarios and detailed commentary when seated; accordingly, we provided seating during the post-game interviews in the main study.

\textit{Darts} constituted a special case. Both the PT co-author and a pilot participant with relevant clinical experience noted its strong resemblance to real-world dart-based training and considered it a natural fit for clinical practice. They suggested that AR glasses could reduce the logistical burden of carrying physical darts in non-fixed testing environments and enable additional game elements to enhance engagement. However, in its current form, \textit{Darts} lacked sufficient technical robustness and usability to serve as a primary choice for our main study set. At the same time, it showed clear potential for redesign into richer PT-aligned training variants, consistent with the forward-looking opportunities identified in our main study. We therefore retained \textit{Darts} as a case study highlighting directions for future redesign rather than including it in the final game set.

\section{Main Study Method}
Our main study aimed to derive design principles for AR Glasses physical-therapy games by treating physical therapists’ \emph{clinical reasoning} as a design lens. Rather than evaluating clinical outcomes, we focused on how PTs—drawing on their routine reasoning—interpret the affordances and limits of AR glasses when engaging with movement-based games. By having therapists experience a curated set of AR games firsthand and reflect on how (and under what conditions) such game mechanics and elements could fit their workflows, we extracted guiding principles and opportunity areas for patient-centered, personalized, and embodied ARPT game design. This section details our study design to achieve these aims.

\subsection{Lightweight AR Glasses - Snap Spectacles}
We conducted the main study on Snap \emph{Spectacles}, a lightweight, standalone, see-through AR glasses platform developed by Snap Inc. We selected Spectacles because they combine an untethered, glasses-class form factor with an on-device stereoscopic display and a comparatively mature community ecosystem of off-the-shelf AR games, enabling us to probe how movement-based play could plausibly align with PT workflows.

\textit{System Configuration.} Spectacles run Snap OS 2.0 and deliver AR experiences as \emph{Lenses}—lightweight application units akin to mobile apps—created in Lens Studio 5. Interaction is primarily embodied: Spectacles support hand/gesture tracking and voice commands with optical and auditory feedback, leveraging computer vision cameras and IMU sensors for 6DoF head and hand tracking. The platform also allows a smartphone as an optional controller. The version used in our study (the Spectacles ’24 developer platform) provides a $\sim$46$^\circ$ diagonal field of view, weighs approximately 226g, and offers about 45 minutes of continuous battery life (extendable via USB-C power), which framed the practical bounds of our play sessions.

Although Spectacles was our study device, we treat it as a proxy for lightweight AR glasses because comparable alternatives were not available at the time: Meta Orion was unreleased, and Xreal relies on external computing and primarily functions as a display.

\subsection{Selected AR Games}
\begin{figure*}[t]
  \centering
  \begin{subfigure}{0.24\textwidth}
    \centering
    \includegraphics[width=\linewidth]{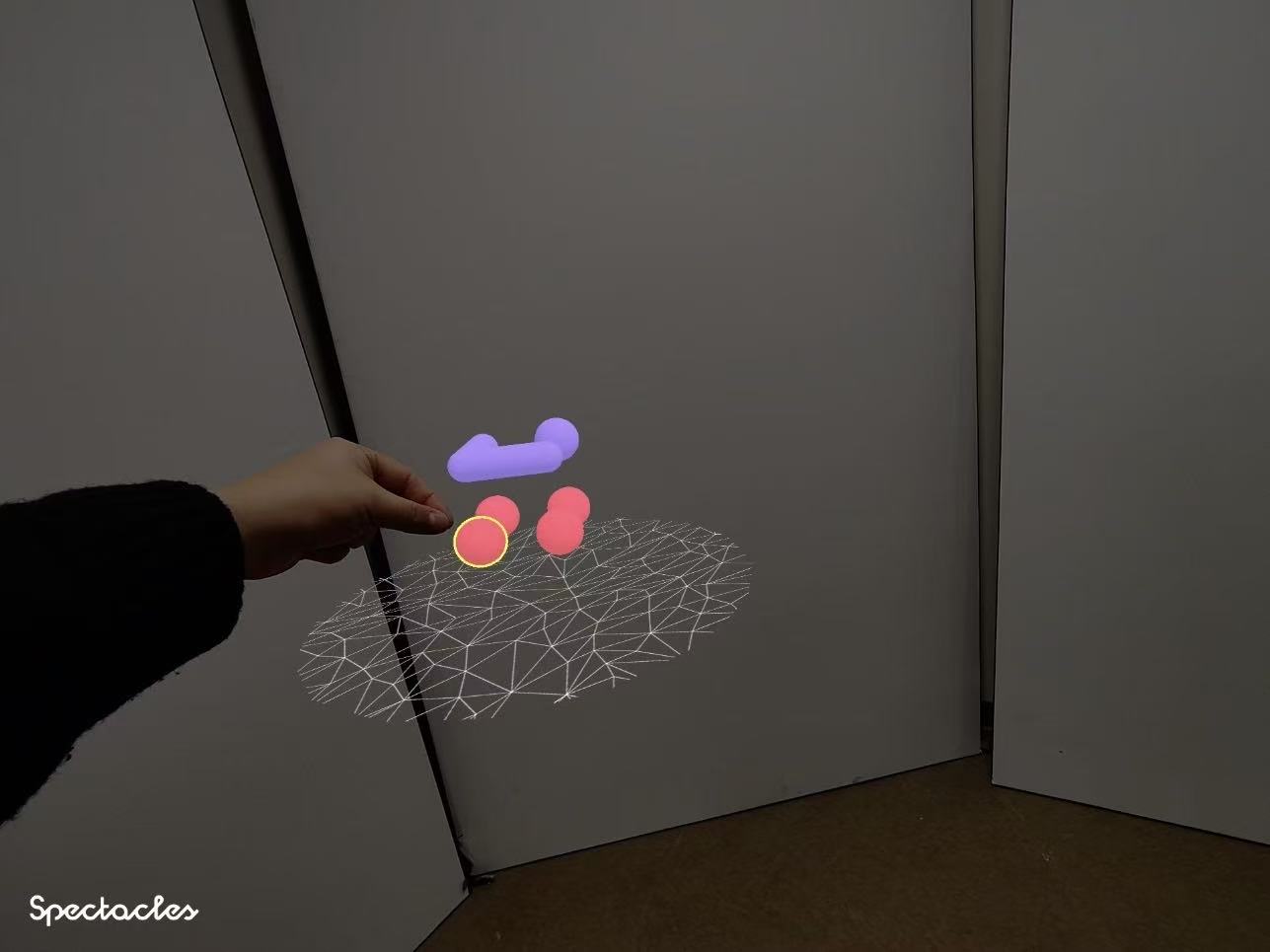} 
    \caption{3-Dots Pinch}
    \label{fig:3dots}
  \end{subfigure}
  \hfill
  \begin{subfigure}{0.24\textwidth}
    \centering
    \includegraphics[width=\linewidth]{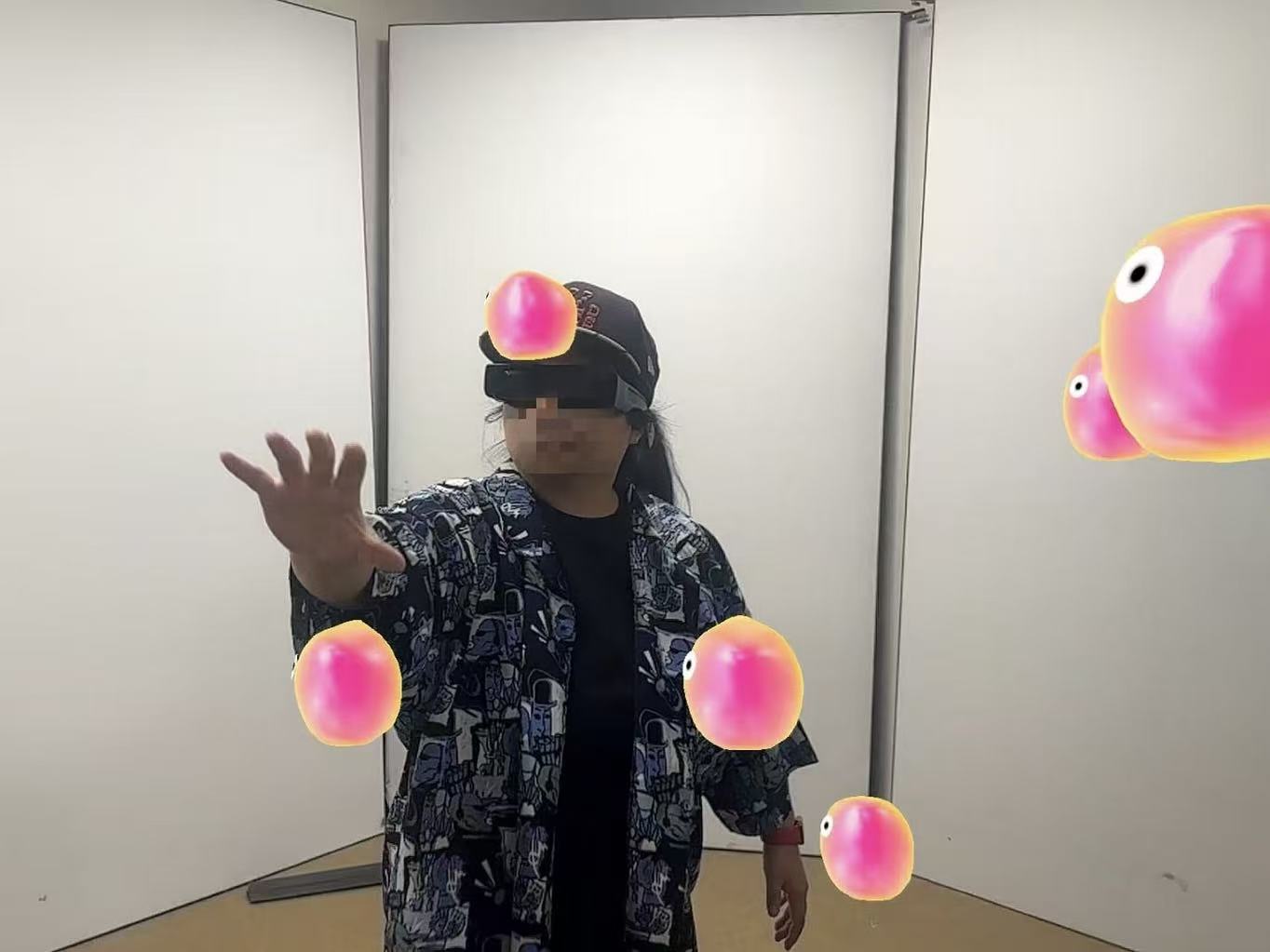} 
    \caption{Squishy Run}
    \label{fig:squishy}
  \end{subfigure}
  \hfill
  \begin{subfigure}{0.24\textwidth}
    \centering
    \includegraphics[width=\linewidth]{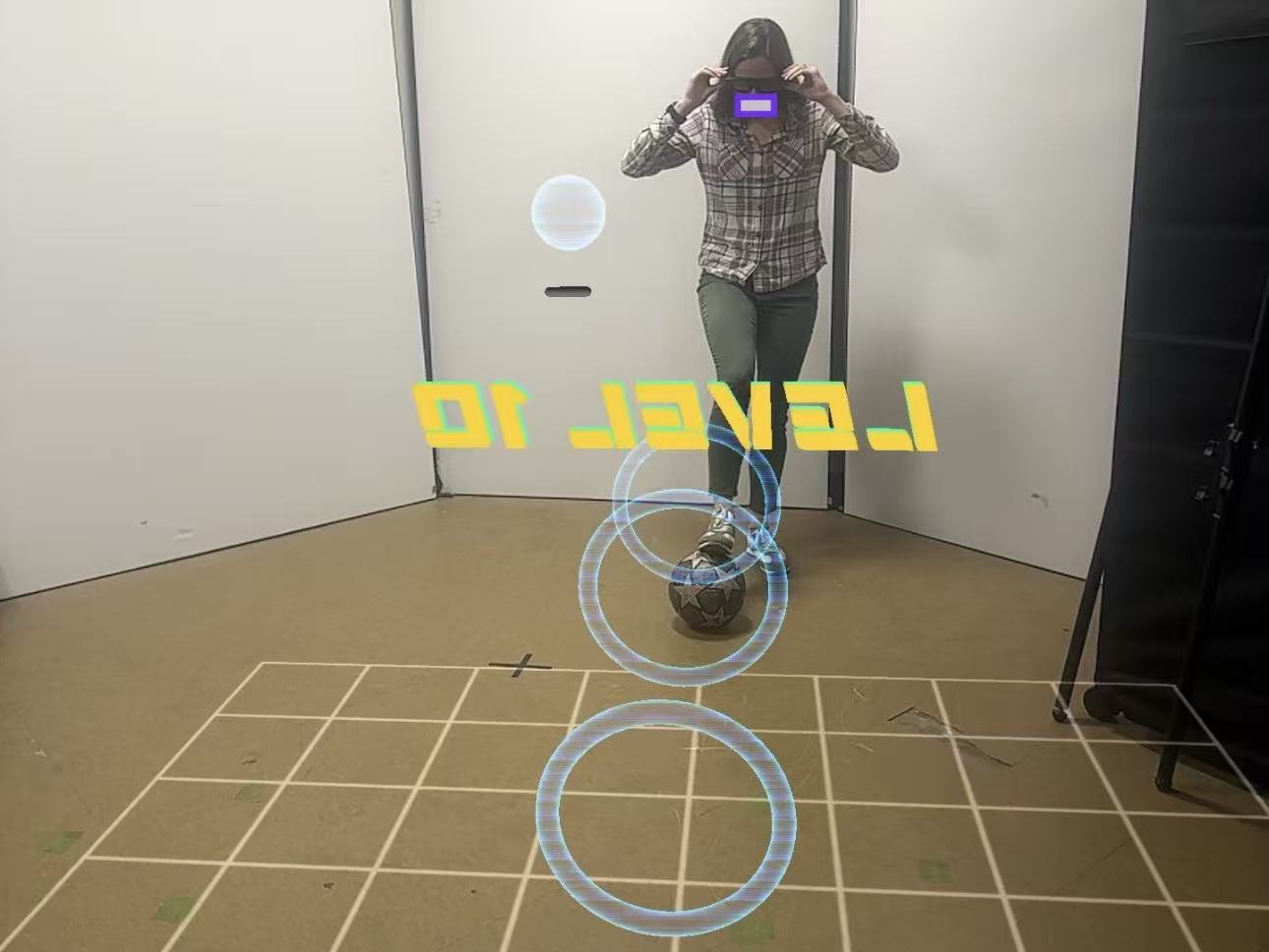} 
    \caption{Ball Game}
    \label{fig:ball}
  \end{subfigure}
  \hfill
  \begin{subfigure}{0.24\textwidth}
    \centering
    \includegraphics[width=\linewidth]{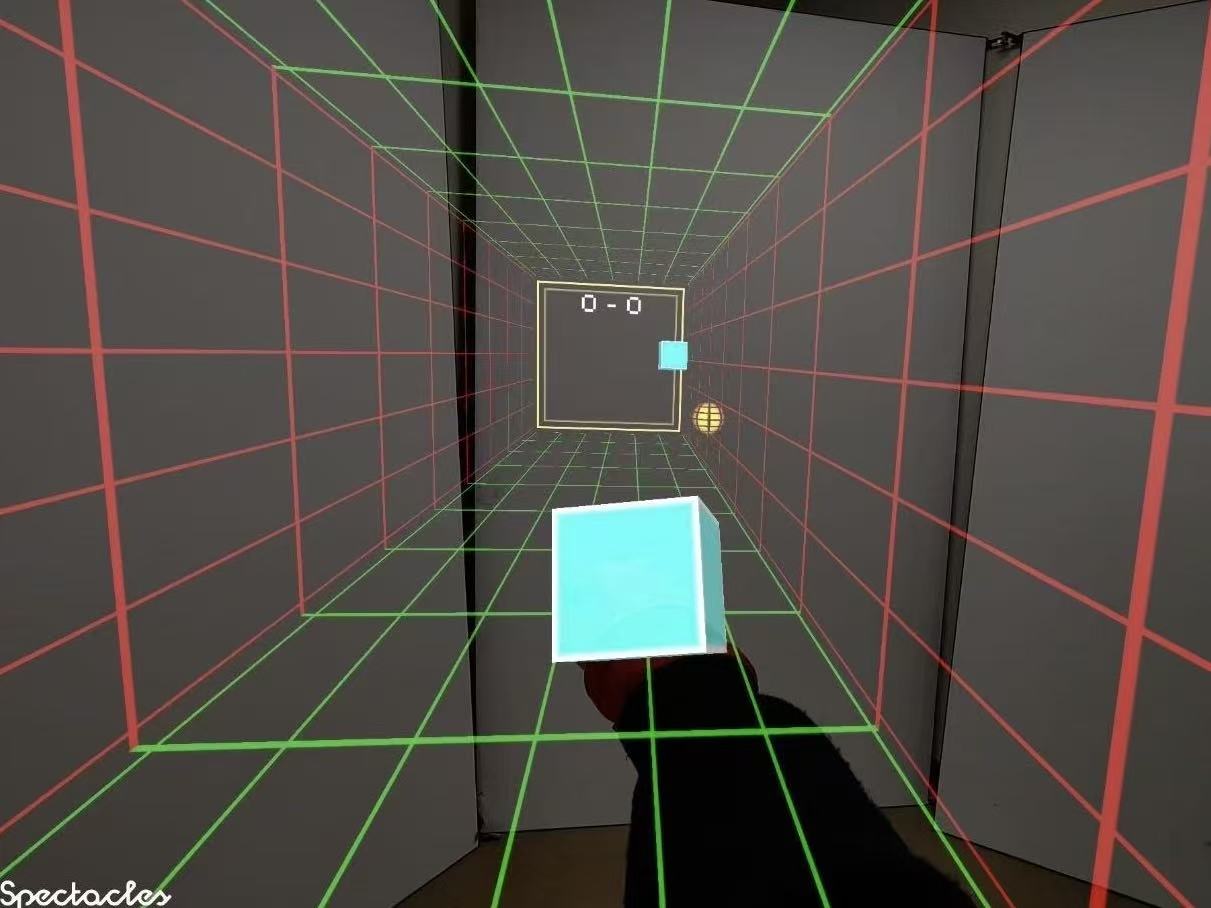} 
    \caption{Action Ball}
    \label{fig:actionball}
  \end{subfigure}

  \caption{
  Four selected Spectacles-based AR games illustrating distinct spatial and motor demands:
  (a) \emph{3-Dots Pinch} integrates fine pinch gestures with whole-body reorientation around 
  a volumetric puzzle;
  (b) \emph{Squishy Run} promotes continuous walking and light upper-limb contact;
  (c) \emph{Ball Game} couples real-ball kicking with virtual moving rings requiring timing;
  (d) \emph{Action Ball} elicits repeated reaching and upper-body adjustments as a virtual ball 
  bounces in a constrained volume.}
  \Description{Composite figure showing four Spectacles-based AR games arranged from left to right. Panel (a) depicts "3-Dots Pinch," a volumetric puzzle controlled through pinch gestures and body repositioning. Panel (b) shows "Squishy Run," which involves continuous walking with intermittent upper-limb contact. Panel (c) presents "Ball Game," combining real-ball kicking with virtual moving targets. Panel (d) illustrates "Action Ball," in which users repeatedly reach and adjust their upper bodies as a virtual ball moves within a bounded space.}

  \label{fig:argames_overview}
\end{figure*}

From the game-selection study, we selected four representative Spectacles-based games (Figure~\ref{fig:argames_overview}).
\textbf{3-Dots Pinch} presents a floating cube populated with colored dots; players use a thumb--index 
pinch gesture to connect dots of the same color while stepping and rotating around the cube to access 
different faces, coupling fine hand control with whole-body reorientation.
\textbf{Squishy Run} distributes soft, blob-like targets (``squishies'') throughout the room; players walk 
through the space to approach each squishy and pat it to make it disappear, promoting continuous 
walking, turning, and light upper-limb contact.
\textbf{Ball Game} combines a tracked physical ball with virtual rings as moving goals; 
players kick the real ball through one or more rings, with later levels adding complex ring motion that 
demands timing and coordination.
\textbf{Action Ball} places a tube-like cubic volume within shoulder reach; a virtual ball bounces inside 
this region, requiring repeated reaching and upper-body adjustments as players strike the ball in 
shifting positions.

\subsection{Participants}
We recruited \textbf{licensed physical therapists (PTs)} as expert participants. All held \textit{Doctor of Physical Therapy (DPT)} degrees and active licensure, trained in the U.S. PT system. PTs integrate biomechanical and physiological knowledge with individualized care, linking movement analysis to therapeutic applications. In clinical practice, PTs work across diverse settings along the care continuum, including hospital-based inpatient and outpatient services, non-hospital outpatient clinics, either general orthopedic or specialty-focused (e.g., sports, neurorehabilitation, pediatric), and home-health care, with most clinicians specializing in one or more areas.

\paragraph{Sampling and procedure.}
We employed \textbf{purposeful maximum-variation sampling}~\cite{Patton2015} to capture diverse reasoning perspectives across specialties (Neuro, Pediatric, Sports, Orthopedics) and career stages (1--37 years). Recruitment occurred through alumni networks, professional contacts, clinic visits, and snowball sampling. Three PTs participated in the pilot (excluded from analysis), and 10 practicing PTs completed the main study. Recruitment continued until thematic saturation was observed~\cite{Guest2006,Malterud2016}.

\paragraph{Professional Profile of PT Participants.}
Table~\ref{tab:demographics} summarizes the professional and clinical characteristics of the ten licensed PT participants. Years of practice ranged from early-career (1 year) to highly experienced clinicians (37 years, mean 14.6). Specialties represented included orthopedics, neurology, pediatrics, sports medicine, and ICU care; roles included travel PT, clinic owners, and high-performance athlete care. Practice settings encompassed outpatient clinics, hospitals (including ICU), and home- and community-based care. Several participants were bilingual (e.g., Spanish, Mandarin, and Korean). This heterogeneity provided a rich cross-section of reasoning perspectives across different patient populations and clinical environments.

\begin{table*}[t]
\centering
\footnotesize
\caption{Professional and clinical characteristics of PT participants (n=10).}
\label{tab:demographics}

\begin{tabularx}{\linewidth}{
  >{\centering\arraybackslash}p{0.8cm}
  >{\centering\arraybackslash}p{2.2cm}
  >{\raggedright\arraybackslash}X
  >{\raggedright\arraybackslash}X
}
\toprule
\textbf{PT ID} & \textbf{Years} & \textbf{Specialty} & \textbf{Primary Setting / Role} \\
\midrule
01 & 37 & Orthopedics; Education & Outpatient PT; PT Professor \\
02 & 1  & Orthopedics & Outpatient PT \\
03 & 1  & Orthopedics; Pediatrics & Outpatient PT \\
04 & 1  & Sports Medicine & Outpatient PT \\
05 & 34 & Neurology; Pediatrics & Hospital PT \\
06 & 1  & Sports Medicine & NBA Player's Personal PT \\
07 & 7  & ICU PT & CVP and AHF care \\
08 & 1  & Orthopedics & Travel PT \\
09 & 36 & Orthopedics & PT Clinic Owner \\
10 & 28 & Orthopedics & PT Clinic Owner \\
\bottomrule
\end{tabularx}
\footnotesize\textit{Note. “Travel PT” = physical therapist working short-term contracts across multiple clinics/regions.}
\end{table*}

\subsection{Experimental Setting}
The study was conducted in two types of environments that mirrored physical therapists’ routine practice. First, sessions took place in a lab in the physical therapy department at Northeastern University, a space routinely used for teaching and research in the Doctor of Physical Therapy (DPT) program. Second, for participants who were unable to travel to the lab, researchers visited their affiliated or owned PT clinics and conducted the study in private treatment rooms, using the equipment and room layout typical of their daily practice. Across both settings, the physical environment closely matched participants’ usual clinical work context.

\subsection{Procedure}
To ground PTs’ feedback in firsthand embodied experience and to elicit their routine \emph{clinical reasoning} in action, we structured each session into three tiers: think-aloud play, post-game interviews, and cross-game reflection. Each session began with a brief orientation. Upon arrival, participants were introduced to the study goals and structure, provided informed consent, and completed a pre-study survey on their clinical PT background and prior use of VR/AR and related technologies (see Supplemental Material). The researcher then introduced Snap Spectacles, demonstrated core functions, and guided participants through a short acclimation period to reduce novelty effects and confirm movement safety. We verified that participants felt comfortable with the device and the planned range of movements, and reiterated that they could pause or withdraw at any time in case of fatigue or discomfort. With consent, all sessions were audio- and video-recorded to preserve natural reasoning and support detailed transcription and analysis.

Participants then completed four game sessions using the final main-study game set, with game order randomized per participant. For each game, participants engaged in a 3-10 minute playtest while thinking aloud, sharing immediate impressions about the gameplay experience, movement demands, and potential relevance to PT practice. Immediately afterward, the researcher conducted a semi-structured interview to probe the game’s perceived clinical meaning, applicability to patient populations, and design considerations. Interview prompts were iteratively refined through the formative pilot and are reported in the next subsection. Throughout the interviews, PTs were encouraged to draw on broader clinical experience beyond the specific games.

Following all four games, participants took part in a short cross-game reflection interview. Having become familiar with the AR glasses by this point, they were asked to compare the games and provide higher-level feedback: (1) which game had the strongest rehabilitation potential and why, (2) which seemed least applicable and why, (3) what key therapeutic elements or capabilities were missing across the set, and (4) what core features they would prioritize in a new AR rehabilitation game. Based on pilot insights showing variability in how PTs preferred to express feedback (open elaboration vs.\ concise responses to direct prompts), this three-tier structure supported both spontaneous and structured articulation of clinical reasoning.

All sessions were conducted in English, lasted approximately 60-120 minutes, and participants were compensated \$20 per hour (not tied to performance or outcomes).

\subsection{Semi-Structured Interview Questions}
Semi-structured prompts were organized into themes aligned with our research aims and iteratively refined through the formative pilot. After each playtest, follow-up questions focused on three per-game themes: \emph{movement relevance}, \emph{engagement and cognition}, and \emph{clinical suitability}. If a participant had already discussed a given issue in detail during the preceding think-aloud, the corresponding prompt was omitted or used only as a brief check for additional comments. After all four games, we held a cross-game reflection interview to elicit comparative judgments and broader design feedback.

\begin{itemize}
    \item \textbf{Movement relevance}
    \begin{itemize}
        \item What movements in the game seem therapeutically useful?
        \item Are these similar to movements you typically use in therapy?
        \item Are there any movements that may be risky or inappropriate for some patients?
    \end{itemize}

    \item \textbf{Engagement and cognition}
    \begin{itemize}
        \item Do you think this game relates to any cognitive functions (e.g., attention, memory, executive function)?
    \end{itemize}

    \item \textbf{Clinical suitability}
    \begin{itemize}
        \item What kinds of patients could benefit from this game?
        \item At which stage(s) of rehabilitation might you use it?
        \item What changes would improve its clinical applicability?
    \end{itemize}

    \item \textbf{Cross-game comparison and general feedback} (asked after all four games)
    \begin{itemize}
        \item Which game has the strongest rehabilitation potential? Why?
        \item Which game seems least applicable? Why?
        \item Are there any key therapy elements missing across these games?
        \item What core features would you prioritize in a new AR rehabilitation game?
    \end{itemize}
\end{itemize}

\subsection{Data Collection and Analysis}
In total, the formative pilot yielded 3.5 hours of video recordings (\emph{excluded from final analysis}), and the main study yielded 12.5 hours of audio/video data across 10 PT sessions (approximately 60-120 minutes per participant). All sessions were recorded, transcribed using Otter.ai, and manually corrected; when relevant to interpretation, transcripts were supplemented with salient nonverbal or embodied cues observed in video (e.g., pauses, laughter, gesture demonstrations, and on-body movement trials). 

We then applied a \textbf{general inductive approach}~\citep{Thomas2006} to derive analytic categories from the data. Two researchers led the qualitative coding. They first collaboratively developed a preliminary codebook on a small subset of transcripts. Next, to calibrate their use of the codebook and assess inter-coder convergence, they independently double-coded 25\% of the transcripts (randomly sampled across participants) and computed Cohen’s $\kappa$, obtaining $\kappa = .68$. After discussing and resolving discrepancies, the researchers independently coded the remaining 75\% of transcripts using the revised codebook. Any subsequent disagreements were resolved through consensus discussions, with a third researcher periodically auditing the evolving codebook and example-coded excerpts.

Low-level codes were iteratively clustered into six analytic domains: \emph{Clinical Alignment (CA)}, \emph{Cognitive \& Perceptual (COG)}, \emph{Risk \& Safety (RS)}, \emph{Technology \& Feedback (TF)}, \emph{Motivation \& Engagement (ME)}, and \emph{Design Principles (DP)}. We normalized all subcodes into these domains; Table~\ref{tab:thematic} and Appendix Table~\ref{tab:crosswalk} report the code--domain crosswalk. We conducted cross-case comparisons by PT specialty, retaining differences rather than collapsing them. Our analytic stance was abductive: categories were grounded in PTs’ narratives while interpreted through clinical reasoning frameworks~\cite{Tavory2014}. This process yielded three higher-order insights—(1) PTs as progressive users, (2) vision--motion differentiation, and (3) patient psychology—which informed ten design guidelines and were synthesized into our conceptual framework (Figure~\ref{fig:model02}).

\begin{figure}[t]
  \centering
  \includegraphics[width=1.1\linewidth]{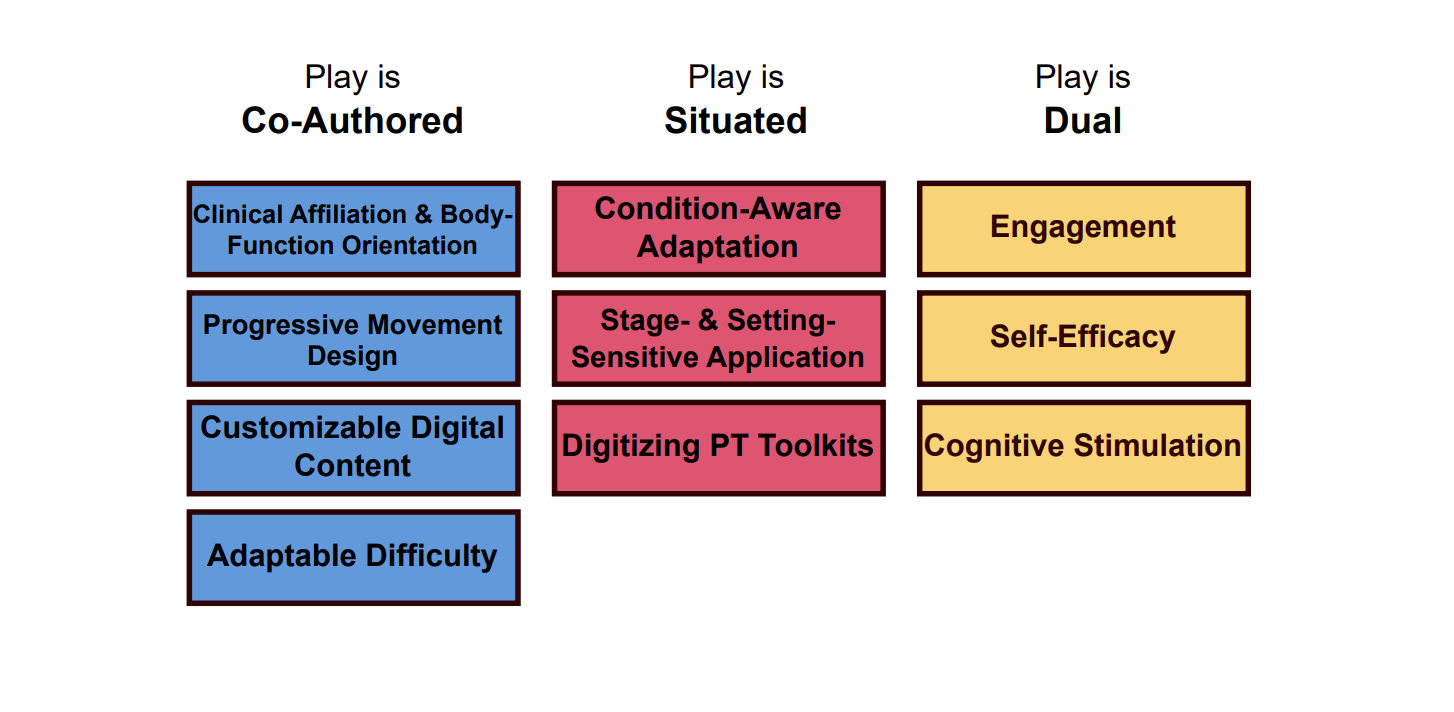}
  \caption{Results of qualitative analysis.}
  \Description{Conceptual diagram summarizing the results of the qualitative analysis. The figure presents three analytic lenses used to organize PT reasoning, with associated subthemes arranged to illustrate relationships among the framework's categories.}

  \label{fig:taxonomy}
\end{figure}

\section{Main Study Results}
Figure~\ref{fig:taxonomy} introduces three lenses for how PTs reason about AR rehabilitation games: \emph{Play is Co-Authored}, \emph{Play is Situated}, and \emph{Play is Dual}. Rather than listing subcategories, we use these lenses to foreground three questions that organized PTs’ design and evaluation judgments: 
(1) \textbf{Who co-authors play, and what are the controllable “knobs”?}—therapist-led parameterization of movements and progressions (e.g., body-region alignment, adjustable targets, non-speed-based difficulty). 
(2) \textbf{For whom and where is play situated?}—condition-, specialty-, and stage-sensitive bounds that change across settings. 
(3) \textbf{What dual effects does play mediate?}—links between physical function and patient psychology (e.g., engagement, self-efficacy, cognitive stimulation). In the sections that follow, each lens is accompanied by a prevalence plot of its subthemes (Figures~\ref{fig:coauthored_codes}, \ref{fig:situated_codes}, and \ref{fig:dual_codes}), and by subcode tables that define the analytic domains and code IDs used in our analysis (Tables~\ref{tab:CA_subcodes}--\ref{tab:CS_subcodes}).

\begin{figure*}[t]
    \centering
    \includegraphics[width=\textwidth]{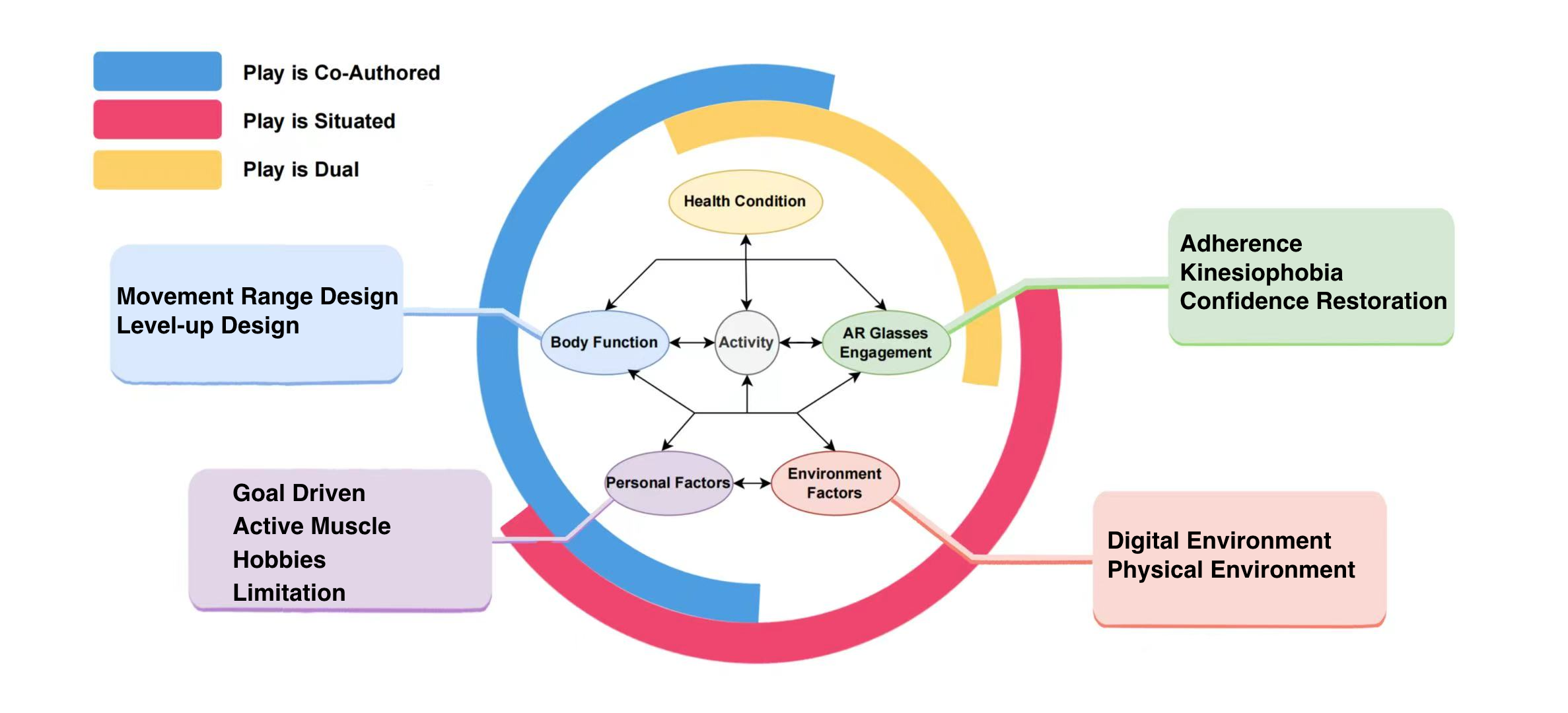}

    \caption{An extension of the ICF framework. While the ICF emphasizes barriers to patient task completion, our model foregrounds PT reasoning about how AR glasses become viable under different conditions (specialty, environment, and patient factors).}
    \Description{Conceptual diagram extending the ICF framework. The figure contrasts the original ICF focus on barriers to task completion with an adapted model that emphasizes physical therapists’ reasoning about when and how AR glasses become viable in practice, incorporating specialty, environmental, and patient-related conditions.}
    \label{fig:model02}
\end{figure*}

These lenses are further synthesized into an extended conceptual model grounded in the ICF framework (Figure~\ref{fig:model02}), which articulates how PT reasoning about AR viability is shaped by clinical, environmental, and patient-level conditions.

Our aim is not to summarize aggregate attitudes toward AR glasses, but to analyze the reasoning processes shaping these judgments. Together, the three lenses and the extended ICF model serve as an analytic scaffold for organizing the empirical themes and PT quotations presented in Table~\ref{tab:thematic}.

\begin{table*}[htbp]

\centering
\caption{Thematic categories, subthemes, and illustrative quotations }
\label{tab:thematic}
\par\small\textit{Domains: CA = Clinical Alignment; COG = Cognitive \& Perceptual; RS = Risk \& Safety; TF = Technology \& Feedback; ME = Motivation \& Engagement; DP = Design Principles.}
\definecolor{coauthored}{RGB}{78,156,255}
\definecolor{situated}{RGB}{239,70,111}
\definecolor{dual}{RGB}{252,208,102}

\small   
\begin{tabularx}{\textwidth}{p{0.18\textwidth} p{0.22\textwidth} X}
\toprule
\textbf{Category (Domain)} & \textbf{Definition} & \textbf{ Example (PT quotes)} \\
\midrule

\multicolumn{3}{l}{\cellcolor{coauthored}\textit{Play is Co-Authored}} \\[0.5ex]
\cmidrule(lr){1-3}

Clinical Affiliation\&Body-Function Orientation (CA) & Movement design aligned with PTs’ body-part logic & “If you program it to reach way up, reach behind, or cross midline, that maps directly to our goals—tight shoulders (ROM/endurance) on the ortho side and midline-crossing for neuro patients after stroke.”\\

Progressive Movement Design (DP) & One game should let PTs scale the same movement from seated → standing → single-leg → dual-task. & “They can play sitting down… [or] standing up… single-leg balance is a higher-level activity… calling out a number adds a dual-task component.”\\

Customizable Digital Content (DP) & PTs can place targets in specific locations to tailor therapy  &  “I want to be able to tell the glasses where to put the targets—put everything on the left so I can force a weight shift; for neglect, place items in the left visual field so the patient has to attend there.”\\

Adaptable Difficulty (TF) & Difficulty adjusted by color, number, size, field-of-view, not just speed/timing   &  “For difficulty, I’d set easy as a smaller reach radius, medium a bit larger, and hard where you reach into all the corners—it doesn’t have to be only about going faster.” \\

\midrule

\multicolumn{3}{l}{\cellcolor{situated}\textit{Play is Situated}} \\[0.5ex]
\cmidrule(lr){1-3}
\\
Condition-Aware Adaptation (CA) & ROM (Range of Motion) must adapt to the patient’s condition.  & “If you had a larger hitting area, you could program the ball to go higher or lower—then the patient would have to use a greater range, and it would challenge them more.” \\

Stage- and Setting-Sensitive Application(RS) & Goals and parameters vary by rehab stage and settings .  & “In the hospital I start with seated balance—patients are really weak—then progress to standing and walking; later a few can use it at home, but only with clear space and safety checks.” \\

Digitizing PT Toolkits (TF) & Virtualizing common PT tools when equipment is limited (ICU, home).  & “In the hospital we don’t have many tools—I’ll blow up a glove for ‘volleyball’ or have them toss socks into a trash can; AR could just be that tool at the bedside.”\\

& Layer digital effects onto existing tools to boost engagement. & “We already put targets on the wall—add the AR lights and sounds and it’s way more motivating than the low-tech version.”\\
\midrule

\multicolumn{3}{l}{\cellcolor{dual}\textit{Play is Dual}} \\[0.5ex]
\cmidrule(lr){1-3}

Engagement (ME) & Games increase enjoyment and adherence  & “Any time you can get a kid doing something that doesn’t feel like therapy is better—you have to gamify PT.”\\

Self-Efficacy (ME) & Wearable AR builds “I can do this” confidence through safe, bite-size wins.  & “For older people who are afraid to move, this gives them something fun to focus on so they start moving again—balance shifts back to being automatic.” \\

& Gradual exposure reduces fear of movement  & “If somebody is afraid of moving, this might be a good distraction to get them moving without really thinking about something being painful or injuring themselves.” \\

Cognitive Stimulation (COG) & Dual-task and sequencing tasks keep the brain engaged & “If there was an ability to call out a number so you had to pick a specific one, that would make it more challenging — it’s adding a dual-task component. Not only do you have to touch it and eliminate it, you now have the additional cognitive task of finding the number first and then doing it.” \\

\bottomrule

\end{tabularx}
\end{table*}

\subsection{AR Rehabilitation Games Framework}

To situate our findings, we propose an \emph{AR Rehabilitation Games Framework}, which adapts the WHO’s 
\emph{International Classification of Functioning, Disability and Health (ICF)}~\cite{WHO2001} to the context of AR game design. 
The ICF is widely used in physical therapy as a scaffold for clinical reasoning—linking body functions and structures, activities and participation, and environmental and personal factors—and as a common language for communication across clinical settings. Extending this role, our adaptation provides a shared vocabulary through which AR developers and clinicians can align design with therapeutic reasoning, embedding digital play directly within practice.

Our adaptation is intentionally modest: we preserve the ICF’s structure while making two targeted changes. First, we replace \emph{Participation} with \emph{AR Glasses Engagement}, clarifying that in our study, the direct users are therapists rather than patients, while still emphasizing how activity-level demands and contextual factors shape what can be done in practice. Second, whereas ICF analysis typically emphasizes environmental or personal barriers to completing tasks, our model foregrounds the \emph{conditionality} of PT adoption—capturing how different specialties view the same AR task through distinct clinical logics (“it depends”). For example, the ICU environment imposes strict sterility and trunk-rotation constraints, making lightweight AR glasses uniquely valuable to fill rehabilitation gaps. More broadly, therapists’ “if–then” reasoning (e.g., “if vestibular issues, then avoid fast visual flow”) anchors our three lenses in a clinically grounded structure that is already usable by PT stakeholders, while opening a space for AR designers to reason about AR game mechanics in ICF terms.

Our analysis revealed three overarching lenses that capture how physical therapists (PTs) reason about AR glasses in rehabilitation: \textit{Play is Co-Authored}, \textit{Play is Situated}, and \textit{Play is Dual}. Each lens comprises several subthemes, described below. These lenses are not merely descriptive; they operationalize how therapists enact embodied interaction and meta-design in practice. When PTs tune movement parameters and progressions, they act as meta-designers who ``keep designing'' AR games in use \cite{FischerGiaccardi2006MetaDesign}. When they shift interpretations of the same game across conditions and environments, they realize Dourish’s view of interaction as situated in concrete practices rather than in abstract interface properties \cite{Dourish2004Context,Dourish2001Action}. Finally, when PTs frame AR play as simultaneously a motor task and a psychological intervention, they highlight rehabilitation games as dual mediators of bodily and experiential change.

\subsection{Play Is Co-Authored}
\begin{figure}[h]
  \centering
  \includegraphics[width=\linewidth]{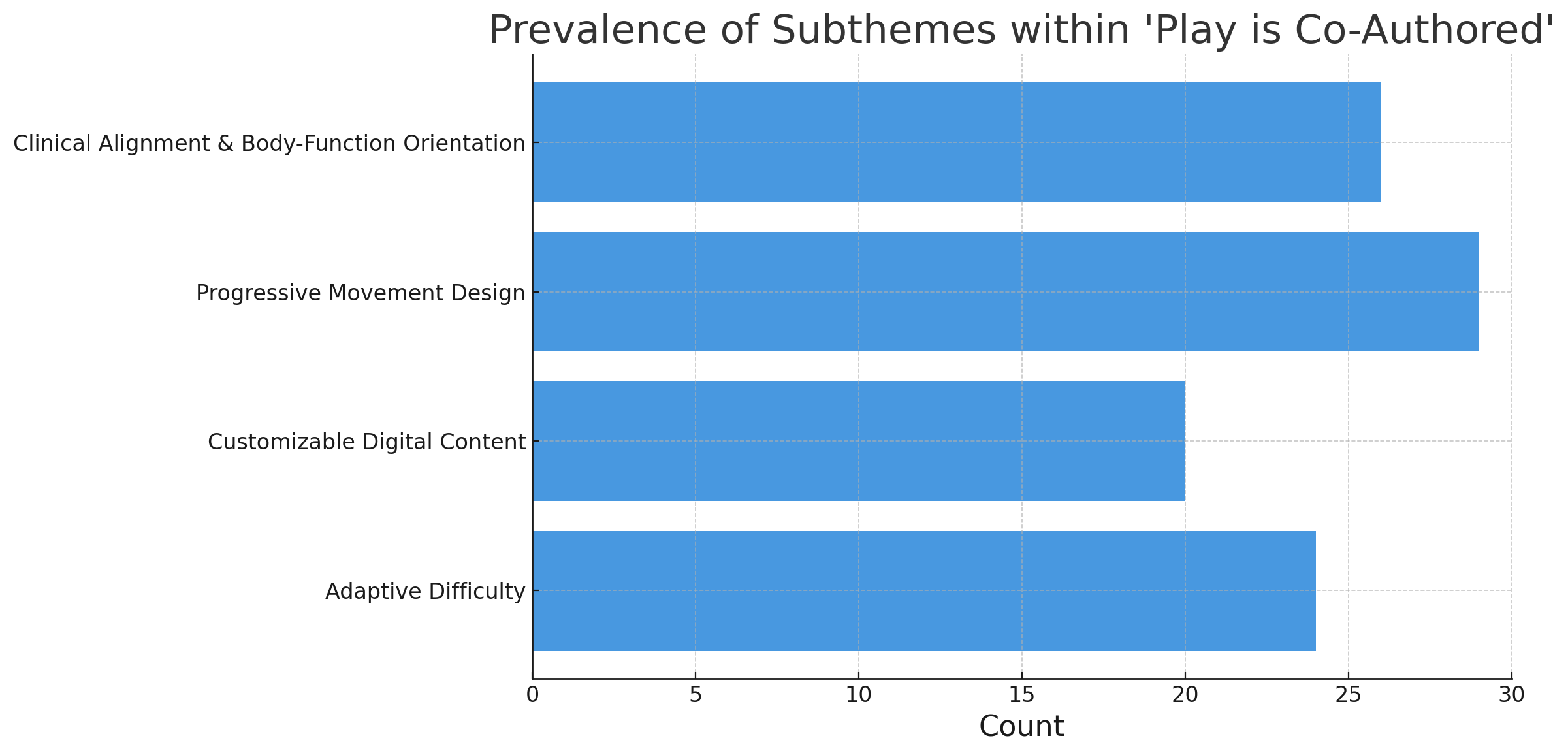}
  \caption{Prevalence of subthemes within \emph{Play is Co-authored}. Bar length indicates the number of coded excerpts per subtheme. Observed coding showed disproportionate emphasis on "Digitizing PT Toolkits", which was normalized in the adjusted framework.}
  \Description{Bar chart showing the prevalence of subthemes within the theme ``Play is Co-authored.'' Each bar represents a subtheme, with bar length corresponding to the number of coded excerpts. The chart visualizes relative differences in coding frequency across subthemes, with one subtheme appearing substantially more frequent than others prior to normalization.}
  \label{fig:coauthored_codes}
\end{figure}

Figure~\ref{fig:coauthored_codes} summarizes the relative prevalence of subthemes within the \emph{Play is Co-Authored} lens across our dataset. To interpret these patterns, we draw on Fischer and Giaccardi’s meta-design theory~\cite{FischerGiaccardi2006MetaDesign}, which positions users as end-users-and-designers: because designers cannot foresee all real-world contingencies, systems should leave a design space that lets users keep designing in use. We extend this view to AR rehabilitation games by framing physical therapists (PTs) as co-authors of play. PTs’ clinical and physiological expertise is essential for re-parameterizing and configuring games to match patients’ diverse conditions and recovery stages; without such co-authorship, AR play has limited clinical reach. Accordingly, we examine the concrete dimensions of the therapist-facing design space—the adjustable “knobs” through which PTs clinically align and progressively tailor AR gameplay.

\subsubsection{Clinical Alignment \& Body-Function Orientation}

A recurring theme in PTs’ accounts was their tendency to anchor AR game interactions in familiar \emph{body-part logics}. Rather than treating embodied interaction mechanics as abstract digital movements, PTs interpreted them through the lens of bodily functions and structures. As one therapist explained while thinking aloud: \emph{This is really about training shoulder flexibility within a certain range\ldots{} of course it also has some cognitive components, like hand--eye coordination} (PT03). Another added: \emph{A single pincer grasp is huge---and holding my arm up, I could feel deltoid fatigue; that's shoulder endurance right there} (PT05).

This orientation reframed even playful mechanics into recognizable therapy categories. For example, what looked like \emph{swatting squishies} was re-authored as \emph{sitting, this is basically trunk control and reaching} (PT07) and, in standing, \emph{it turns into a balance and spatial-scanning task} (PT03). Through such mappings, PTs translated ambiguous digital actions into concrete rehabilitation practices and began placing them on progression ladders. At the same time, this orientation acted as a filter: if an AR task could not be clearly mapped to a body region or functional outcome, PTs often down-weighted the task's clinical relevance.

Our codebook analysis clarifies that \emph{Clinical Alignment} itself decomposes into four subcategories (Table~\ref{tab:CA_subcodes}):  
\begin{table}[!htbp]
\centering
\caption{Subcodes within the Clinical Alignment domain.}
\label{tab:CA_subcodes}
\definecolor{coauthor1}{RGB}{78,156,255}
\par\small\textit{Domain: CA = Clinical Alignment. CodeIDs here use the \textbf{CA-} prefix and full names (no abbreviations); see Appendix Table~\ref{tab:crosswalk} for the code–domain crosswalk.}
\begin{tabularx}{0.9\linewidth}{l X}
\toprule
\rowcolor{coauthor1}
\textbf{CodeID} & \textbf{Definition} \\
\midrule
CA-FunctionalOutcomes      & Mapping to functional outcomes (e.g., range of motion, balance). \\
CA-ActivitiesOfDailyLiving & Links to activities of daily living (e.g., walking, reaching). \\
CA-PatientPopulations      & Alignment with specific patient populations (e.g., pediatrics, stroke). \\
CA-BodyRegion              & Explicit association with body regions (shoulder, trunk, lower limb). \\
\bottomrule
\end{tabularx}
\end{table}

In this sense, clinical alignment constitutes the first step of co-authorship:  PTs position themselves as interpreters who translate digital play into the concrete vocabulary of rehabilitation practice, ensuring that ambiguous AR interactions are reframed into actionable clinical categories.

\subsubsection{Progressive Movement Design}
PTs consistently emphasized that AR rehabilitation games must support \emph{graded progression}, echoing principles of physical therapy. They described progression not only in terms of task difficulty, but also as structured transitions between postures, ranges, and multi-task challenges. As one PT emphasized posture options: \emph{They can play sitting down if they need to\ldots{} I think it's safer because they have different options} (PT02). Another, who works on an inpatient unit, added: \emph{I could see working on sitting balance\ldots{} with my patients who are really, really sick\ldots{} build your core strength and work on reaching, even in sitting} (PT07). In this framing, digital games are not one-off activities, but scaffolds that should grow with the patient’s recovery trajectory.

A critical part of this progression occurs even \emph{before} gameplay begins. Therapists stressed that clinical reasoning requires assessing the patient’s current limitations and calibrating digital tasks accordingly. For example, when treating a frozen shoulder patient with only 30° of abduction, a PT explained: \emph{I would set the target range lower from the start---otherwise the game is asking for something the patient simply cannot do} (PT04). Such pre-game calibration ensures that digital progression is anchored in realistic therapeutic ranges, transforming an otherwise generic task into a clinically viable one.

Progression also required \emph{in-play adaptation}. Therapists described watching for patterns of asymmetry or neglect-for example, patients growing overly comfortable with one limb or one side of space while consistently ignoring the other. In such cases, PTs adjusted the frequency, position, or size of digital targets to draw attention back to the underused area (e.g., in post-stroke hemiparesis/hemiplegia, encouraging engagement of the affected side to support motor relearning and neuroplasticity): \emph{If I notice the patient keeps hitting everything on the right and skipping the left, I’ll increase the targets on the left side until they are forced to engage it} (PT09). This ability to modify tasks on the fly reinforced the view that AR systems should remain open-ended, clinician-configurable tools rather than pre-scripted, non-configurable experiences.

Progression was further imagined along multiple axes—posture (seated $\rightarrow$ standing $\rightarrow$ single-leg), coordination complexity (single gesture $\rightarrow$ sequencing $\rightarrow$ dual-task), and stability demands (stable surface $\rightarrow$ balance challenges). Therapists also emphasized micro-progression within a single session: \emph{You can tailor it to the clinic—easy reach here, then higher, behind you, cross midline} (PT04). A pediatric PT described increasing the challenge of a balloon-popping task by switching hand dominance, then requiring cross-body reaches, and finally adding stepping movements (PT05). Therapists further advocated mixing motor with cognitive layers when appropriate.

Failure to accommodate such gradation, however, was viewed as a major barrier to clinical adoption. Games that increased difficulty exclusively via speed were frequently described as frustrating or unsafe, particularly for early-stage patients. One therapist put it bluntly: \emph{If the only way it gets harder is by going faster, that's not progression---it's just punishment} (PT01). Another noted: \emph{I don't think that one ring to two rings to three rings is how I would go up the levels\ldots{} I'd go one ring \textit{slow}, then one ring \textit{fast}} (PT02). In line with this, PTs asked for independent difficulty levers (e.g., reach range, radius, target count/density, rule complexity) to support gradual motor learning without undermining confidence.

Our coding further broke down progressive movement design into five recurring emphases (Table~\ref{tab:PMD_subcodes}):  

\begin{table}[!htbp]
\centering
\caption{Subcodes within the Progressive Movement Design (PMD) domain. 
PMD = Progressive Movement Design.}
\label{tab:PMD_subcodes}
\definecolor{coauthor1}{RGB}{78,156,255}
\par\small\textit{Domain: DP = Design Principles. CodeIDs here use the \textbf{PMD-} prefix; see Appendix Table~\ref{tab:crosswalk} for the code–domain crosswalk.}
\begin{tabularx}{0.9\linewidth}{l X}
\toprule
\rowcolor{coauthor1}
\textbf{CodeID} & \textbf{Definition} \\
\midrule
PMD-PreCalibration & Pre-game adaptation based on clinical reasoning (e.g., setting target range to fit frozen shoulder limits). \\
PMD-InPlayAdaptation & On-the-fly adjustment of object frequency, position, or size to address asymmetries or neglect (e.g., increasing targets on the left side). \\
PMD-Posture & Scaling across seated, standing, single-leg, and dynamic balance tasks. \\
PMD-Complexity & Increasing coordination demands (sequencing, dual-task, cross-body). \\
PMD-Load & Adjustments through resistance, instability, or surface (e.g., balance board). \\
\bottomrule
\end{tabularx}
\end{table}

In this sense, progressive movement design underscores PTs' co-authorship: AR experiences are judged not as static tasks but as \emph{dynamic progressions}. These begin with pre-session calibration, continue with in-play adaptation, and extend across postural, coordinative, and stability dimensions-mirroring therapeutic trajectories and giving PTs flexible levers for intensity, stability, and complexity.

\subsubsection{Customizable Digital Content}

As noted in the previous section on \emph{in-play adaptation}, therapists emphasized that AR games must provide \emph{customizable digital content} tailored to patients’ specific goals and impairments. Unlike traditional rehabilitation games, which often offer fixed targets or preset difficulty levels, PTs envisioned AR glasses as a flexible toolkit in which digital elements could be placed, resized, or sequenced according to clinical logic. One therapist explained that \emph{Sometimes I want the targets only on the patient’s left side, to train neglect} (PT05), while another said that \emph{sometimes I want them positioned higher and closer to encourage shoulder flexion without strain} (PT02).

Customization was seen not merely as a convenience but as foundational to aligning digital play with clinical reasoning. Therapists highlighted that patient conditions often require selective emphasis—for example, increasing the frequency of targets for a weaker limb, constraining the task to safe ranges, or scaling stimulus size to bolster confidence. As they noted, the system should not dictate what matters; rather, parameters must be adjustable so therapists can emphasize what they observe in real time: \emph{I don’t want the game to decide what’s important; I want to decide based on what I see in front of me} (PT09).

Several therapists explicitly drew analogies to their everyday use of physical props. Just as cones, resistance bands, or balance boards are repositioned to meet individual needs, they imagined AR games as digital extensions of these tools. In this sense, customizable content was not an “extra” feature, but a necessary condition for AR systems to function as clinically meaningful instruments.

\begin{table}[h]
\centering
\caption{Subcodes within the Customizable Digital Content (CDC) domain. 
CDC = Customizable Digital Content.}
\label{tab:CDC_subcodes}
\definecolor{coauthor1}{RGB}{78,156,255}
\par\small\textit{Domain: DP = Design Principles. CodeIDs here use the \textbf{CDC-} prefix; see Appendix Table~\ref{tab:crosswalk} for the code–domain crosswalk.}
\begin{tabularx}{0.9\linewidth}{l X}
\toprule
\rowcolor{coauthor1}
\textbf{CodeID} & \textbf{Definition} \\
\midrule
CDC-Target Placement & Selectively placing digital targets to train specific deficits (e.g., left neglect, limb weakness). \\
CDC-Stimulus Scaling & Adjusting frequency, size, color, or salience of stimuli to scaffold difficulty and confidence. \\
CDC-Therapist Control & Preserving therapist authority to define therapeutic priorities in real time. \\
\bottomrule
\end{tabularx}
\end{table}

Several therapists cautioned that fixed targets or presets may limit clinical usefulness. As one remarked: ``If I can’t adapt it to the patient, then it’s just a game, not therapy'' (PT06). While not universal, this view suggests that adjustable parameters can enable PTs to integrate AR tasks into existing practice rather than treating them as stand-alone games.

\subsubsection{Adaptable Difficulty}

Complementing Sections~4.2.2 and 4.2.3, therapists emphasized the need for \emph{Adaptable Difficulty} at the level of game systems.
Whereas progressive movement design focuses on sequencing therapeutic actions across stages, and customizable content ensures therapist control over \emph{what} is emphasized, Adaptable Difficulty determines \emph{how} difficulty scales in play.
Therapists criticized speed- or repetition-only models as reductive and demotivating: \emph{I don’t want the game to level up just by being faster—that frustrates patients} (PT08).
They also flagged timing-only escalation as clinically misaligned (e.g., a soccer level that requires synchronous kicks through two shifting goals): \emph{This game point is really about timing—it doesn’t actually connect to the body at all} (PT02).

Instead, PTs described difficulty as inherently \emph{multi-dimensional}.
Progression can be achieved by manipulating (1) \textbf{target properties} (size, color, number, salience), (2) the \textbf{spatial envelope} or field-of-view (e.g., higher reaches, lateral or cross-body placements), and (3) \textbf{temporal patterns} beyond raw speed (e.g., variable intervals, predictable rhythms that preserve motor control).
They also advocated \textbf{rule-based overlays} (sequencing, memory, selective attention) to increase cognitive load without unsafe kinetic demands. PT05 noted that introducing features such as calling out specific numbers for the user to select would increase task difficulty by adding a dual-task component; similarly, \emph{It's a nice way to do dual tasking\ldots{} balance while making color associations} (PT01).

Safety coupling was a recurring concern. Therapists emphasized bounded ranges and environment setup when difficulty pushes reach or movement (\emph{I would set the target range lower from the start\ldots{} otherwise the game is asking for something the patient simply cannot do} (PT04); \emph{It would have to be done in a very clear, open space\ldots{}} (PT08)) and highlighted supportive feedback as difficulty increases. Importantly, these adjustments were framed as \emph{therapist-facing levers}, not fixed game logics, so that challenge can be tuned without eroding confidence or safety. Several therapists also requested a \emph{preview/test mode} to trial parameter changes before handing glasses back to the patient, ensuring that escalation remains clinically appropriate.
\begin{table}[h]
\centering
\caption{Subcodes within the Adaptable Difficulty (AD) domain. 
AD =  Adaptable Difficulty.}
\label{tab:AD_subcodes}
\definecolor{coauthor1}{RGB}{78,156,255}
\par\small\textit{Domain: TF = Technology \& Feedback. CodeIDs here use the \textbf{AD-} prefix; see Appendix Table~\ref{tab:crosswalk} for the code–domain crosswalk.}
\begin{tabularx}{0.9\linewidth}{l X}
\toprule
\rowcolor{coauthor1}
\textbf{CodeID} & \textbf{Definition} \\
\midrule
AD-TargetParams & Scale difficulty via target number, size, color, salience, or spawn rate. \\
AD-SpatialEnvelope & Expand/shift the field-of-view (higher reach, lateral/cross-body zones) within safe bounds. \\
AD-TemporalPattern & Adjust rhythms and intervals (not just speed) to preserve control and pacing. \\
AD-CognitiveLayering & Add sequencing, memory, or attention rules to raise complexity without unsafe load. \\
AD-SafetyAssist & Guardrails and assistive tuning (hitbox enlargement, “magnetism,” auto-ease after failures). \\
\bottomrule
\end{tabularx}
\end{table}

In short, adaptable progression mechanics make co-authorship actionable at the micro level: by exposing multidimensional scaling within the system, AR games can align moment-to-moment difficulty with therapeutic trajectories, rather than forcing one-size-fits-all escalation.

\subsection{Play Is situated}
\begin{figure}[H]
  \centering
  \includegraphics[width=\linewidth]{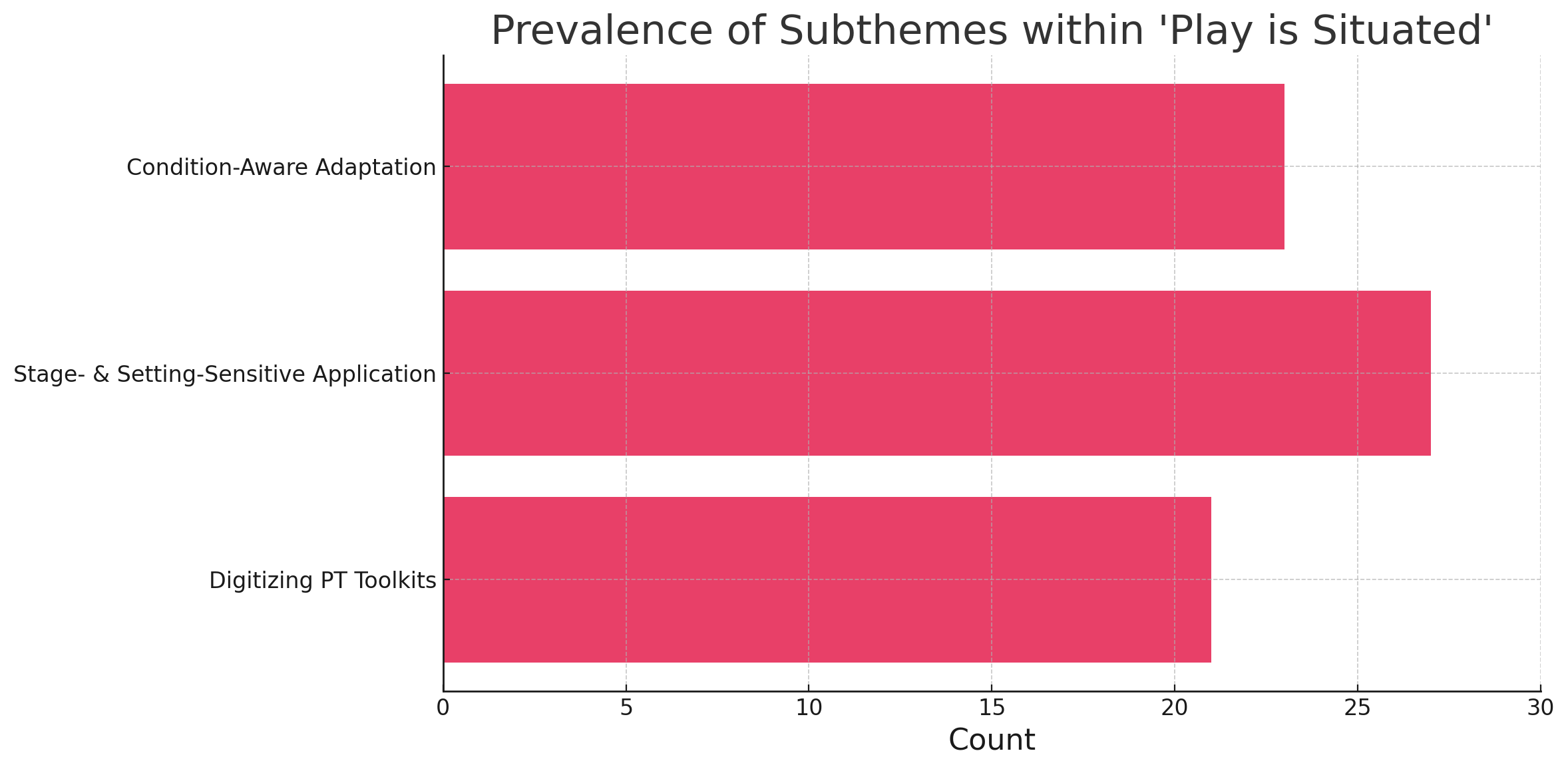}
  \caption{Prevalence of subthemes within \emph{Play is Situated}. Bar length indicates the number of coded excerpts per subtheme.}
  \Description{Bar chart showing the prevalence of subthemes within the theme "Play is Situated." Each bar represents a subtheme, with bar length corresponding to the number of coded excerpts, allowing comparison of relative coding frequency across context- and setting-related subthemes.}

  \label{fig:situated_codes}
\end{figure}
Figure~\ref{fig:situated_codes} shows how often each situated subtheme was coded in therapists' accounts. We ground this theme in Dourish’s theory of embodied interaction~\cite{Dourish2004Context,Dourish2001Action}, which argues that interaction gains meaning through situated bodily action rather than abstract input–output mappings. This lens underpins our notion of situated play: AR motion-based gameplay is not simply a set of motor-control tasks, but an embodied practice configured by patients’ perceptions, therapeutic movement logics, and the concrete environments of hospitals, clinics, and homes. In our data, therapists consistently treated AR game elements as augmentations of real-world tools and spatial cues, reinforcing that play is always interpreted through specific bodies, conditions, and settings.
\subsubsection{Condition-Aware Adaptation}

Therapists repeatedly emphasized that AR rehabilitation games must be adaptable to each patient's specific conditions and limitations. A central theme was that one-size-fits-all mechanics often clash with the realities of therapy: \emph{Frozen shoulder patients have very different limits. If the game sets the same range for everyone, those with restrictions get frustrated and disengage} (PT02). Another therapist contrasted this with Parkinson’s therapy: 
\emph{The LSVT BIG program trains them to move big—so a game with forceful, exaggerated punches is actually useful for overcoming their tendency toward small, slow motions} (PT08; see also~\citep{Peterka2020LSVT}). For stroke or arthritis patients, the concern was even more basic: \emph{They might never reach 90 degrees. If that’s the minimum required, the system excludes them entirely} (PT07).

This logic extended beyond musculoskeletal impairments to neurological and visual conditions. For instance, one therapist described working with stroke survivors who had lost central vision and relied on deliberate head movements to scan. Traditional ribbon or cone exercises felt punitive and demotivating. In contrast, AR glasses were envisioned as a medium to \emph{augment visual space}, dynamically placing digital targets in the periphery to scaffold scanning in more engaging ways. Another added, \emph{Patients with dysmetria misjudge distance and trip off curbs. Games that train spatial judgment and foot-eye coordination could really help} (PT05).

Such reasoning underscores that condition-specific adaptation is not merely about adjusting movement ranges but about tailoring digital play to each patient’s embodied constraints and capacities. As one PT warned, \emph{If there are no constraints, patients will just walk instead of reaching. Compensation replaces the actual therapy} (PT03). AR was seen as valuable not only for its adaptability, but for its potential to transform impairments into structured, meaningful challenges—when designed with full-body and full-context flexibility in mind.
\begin{table}[h]
\centering
\caption{Subcodes within the Condition-Aware Adaptation (CAA) domain. 
CAA = Condition-Aware Adaptation.}
\label{tab:CAA_subcodes}
\definecolor{situated1}{RGB}{239,70,111}
\par\small\textit{Domain: CA = Clinical Alignment. CodeIDs here use the \textbf{CAA-} prefix; see Appendix Table~\ref{tab:crosswalk} for the code–domain crosswalk.}
\begin{tabularx}{0.9\linewidth}{l X}
\toprule
\rowcolor{situated1}
\textbf{CodeID} & \textbf{Definition} \\
\midrule
CAA-RangeCalibration & Adjusting digital task ranges to fit condition-specific movement limits (e.g., frozen shoulder). \\
CAA-JointSafety & Ensuring posture and loading safety for specific joints (e.g., cervical spine, weight-bearing). \\
CAA-AsymmetryCompensation & Increasing target frequency/placement on weaker or neglected sides. \\
\bottomrule
\end{tabularx}
\end{table}

\subsubsection{Stage- \& Setting-Sensitive Application}

PTs also stressed that the applicability of AR rehabilitation games depends strongly on the rehabilitation stage and the treatment setting. Early-stage patients, such as those recovering in intensive care units (ICUs), often have limited mobility, require close guarding, and face strict sterility requirements. As one therapist explained: \emph{In the ICU, the patient is in a critical care bed, and I need to guard trunk rotation closely. I can’t set them tasks that involve standing or walking just yet} (PT07). In such contexts, AR glasses were envisioned as enabling low-intensity seated exercises, carefully bounded to avoid fatigue or risk.

By contrast, in later outpatient or sports rehabilitation stages, therapists expected AR games to scale up toward higher intensity and greater complexity. For athletes, this meant progression to agility or endurance drills; for pediatric patients, incorporating gamified challenges that sustain motivation; and for neurological rehabilitation, embedding dual-task demands that combine motor actions with cognitive stimulation. As PT06 described: \emph{For athletes, I need soccer-like drills—something that actually feels like their sport.}
\begin{table}[!htbp]
\centering
\caption{Subcodes within the Stage- and Setting-Sensitive Application domain.}
\label{tab:SSSA_subcodes}
\definecolor{situated1}{RGB}{239,70,111}
\par\small\textit{Domain: RS = Risk \& Safety. CodeIDs here use the \textbf{SSSA-} prefix; see Appendix Table~\ref{tab:crosswalk} for the code–domain crosswalk.}
\begin{tabularx}{0.9\linewidth}{l X}
\toprule
\rowcolor{situated1}
\textbf{CodeID} & \textbf{Definition} \\
\midrule
SSSA-EarlyStage   & Adapting AR tasks for low-intensity, seated, or closely guarded use in acute/ICU settings. \\
SSSA-LateStage    & Scaling AR games for higher intensity, agility, or dual-task demands in outpatient, sports, or neuro rehab. \\
SSSA-Environment  & Tailoring exercises to physical setting constraints (e.g., small home spaces, hygiene in shared clinics). \\
\bottomrule
\end{tabularx}
\end{table}

The environmental setting further shaped reasoning. At home, space constraints and safety risks (e.g., furniture, obstacles, or lack of supervision) limited the types of movements considered feasible. One therapist warned: \emph{At home, even a rug can trip them. If the game expects them to sidestep or spin, that’s dangerous without someone there} (PT06). Another added: \emph{I can’t send patients home with a system that doesn’t adjust. If it forces them into actions beyond their safety, I get liability concerns—and they’ll just stop using it} (PT08). In contrast, clinics allowed for more ambitious full-body tasks, but also raised concerns about hygiene when sharing equipment. Therapists thus imagined AR glasses as providing context-sensitive flexibility—able to deliver simple, contained exercises in constrained spaces, while scaling up to more demanding activities in supervised clinics.

Our coding clustered these perspectives into three subcodes (Table~\ref{tab:SSSA_subcodes}). Together, they highlight how therapists link AR rehabilitation games to both temporal stages of recovery and the affordances or constraints of treatment environments.

\subsubsection{Digitizing PT Toolkits}

Therapists frequently drew analogies between AR glasses and their existing arsenal of physical tools. Balls, cones, wall targets, elastic bands, and
balance surfaces are staples of everyday rehabilitation, valued for
familiarity, flexible setup, and graded progression. Participants consistently imagined AR as a way to \emph{digitally extend} these toolkits—offering virtual versions deployable across settings. As one PT put it, \emph{“Kicking we use a lot to affect a weight shift—single-leg support balance. There are a lot of reasons we use balls in the clinic”} (PT04).

This digital extension was framed as complementary rather than a replacement— especially in constrained contexts (ICU, inpatient wards, home) where equipment is limited or must be sterilizable. As one inpatient PT explained, \emph{“We don’t have many tools—in the hospital I’ll blow up a glove for ‘volleyball’ or have them toss socks into a trash can”} (PT07). In such settings, AR can virtualize common props (e.g., target cones, wall dots, moving cues) while preserving clinical logics (repetition, range, balance challenge) and adding multimodal feedback, adaptable scaling, and on-the-spot reconfiguration.

Beyond digitizing familiar clinic props, therapists emphasized that AR glasses could also \emph{augment} the physical tools they already rely on. Because Spectacles and emerging AR glasses support physical–object tracking, PTs imagined hybrids where traditional equipment—balls, cones, therapy bands, foam pads—becomes an interactive anchor for adaptable cues, spatial constraints, or graded challenges. Rather than replacing real objects, AR can layer virtual targets, trajectories, or timing indicators onto them, enabling richer feedback while preserving the tactile qualities essential to rehabilitation. As one PT noted, such hybrid tools could \emph{make the tools we already trust do more work for us} (PT03)—turning everyday equipment into extensible platforms for progression and context-specific modification.
\begin{table}[h]
\centering
\caption{Subcodes within the Digitizing PT Toolkits (DPT) domain. 
DPT = Digitizing PT Toolkits.}
\label{tab:DPT_subcodes}
\definecolor{situated1}{RGB}{239,70,111}
\par\small\textit{Domain: TF = Technology \& Feedback. CodeIDs here use the \textbf{DPT-} prefix; see Appendix Table~\ref{tab:crosswalk} for the code–domain crosswalk.}
\begin{tabularx}{0.9\linewidth}{l X}
\toprule
\rowcolor{situated1}
\textbf{CodeID} & \textbf{Definition} \\
\midrule
DPT-VirtualizedProps & Familiar PT tools (balls, cones, bands) re-created as interactive digital elements. \\
DPT-AllInOne & AR substitutes for physical tools in constrained contexts (ICU, home, travel), reducing equipment burden. \\
DPT-AugmentedFeedback & Digital props enriched with visual, auditory, or adaptive feedback unavailable in physical form. \\
\bottomrule
\end{tabularx}
\end{table}
Our coding clustered these perspectives into three subcodes (Table~\ref{tab:DPT_subcodes}). 
Together, they highlight how digital toolkits ground AR rehabilitation in therapists’ established practice while opening space for new forms of progression, feedback, and contextual adaptability.

In summary, digitizing PT toolkits reflects therapists’ desire for continuity: new technology should not impose unfamiliar logics, but rather extend and enrich the materials they already trust. 
Most PTs had no prior hands-on experience with AR glasses, and early remarks were mixed, balancing excitement and caution; as sessions progressed, their talk shifted from broad speculation to concrete, clinically grounded comparisons across exemplar tasks (see Methods).
This theme captures how participants translated a novel medium into familiar rehabilitation reasoning—treating AR elements as placeable, resizable, and sequenceable ``digital props'' that can be woven into progression ladders and dose control—thus foregrounding co-authorship in day-to-day practice.

Taken together, these subthemes show that AR rehabilitation games are inherently \emph{situated}: their meaning and feasibility shift with conditions, joints, recovery stages, and treatment settings. What may appear as universal mechanics (e.g., kicking a ball, reaching a target) are re-authored through distinct clinical logics—whether calibrating ranges for a frozen shoulder, embedding scanning tasks for vision loss, or adapting exercises for the sterility of ICUs versus the constraints of home care. This situatedness is not a limitation but a design principle, underscoring the need for adaptable systems that bridge laboratory prototypes and diverse clinical realities.

\subsection{Play Is Dual}
\begin{figure}[htbp]
  \centering
  \includegraphics[width=\linewidth]{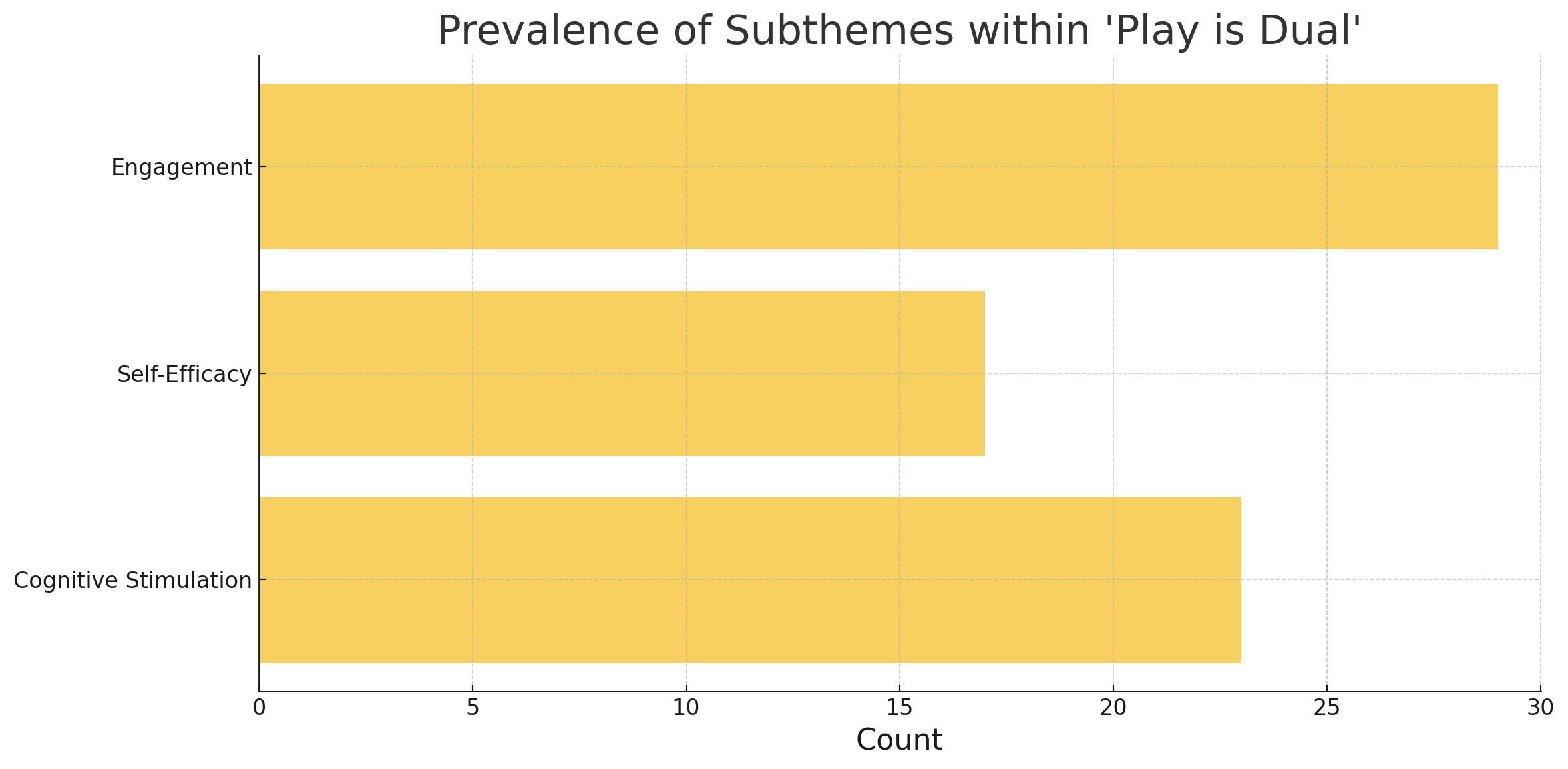}
  \caption{Prevalence of subthemes within \emph{Play is Dual}. Bar length indicates the number of coded excerpts per subtheme.}
  \Description{Bar chart showing the prevalence of subthemes within the theme "Play is Dual." Each bar represents a subtheme, with bar length corresponding to the number of coded excerpts, enabling comparison of relative emphasis across psychological and functional dimensions of play.}

  \label{fig:dual_codes}
\end{figure}
Figure~\ref{fig:dual_codes} visualizes the distribution of codes within the \emph{Play is Dual} lens. 
We ground this theme in Self-Determination Theory (SDT), proposed by Deci and Ryan~\cite{DeciRyan1985SDT},
one of the most widely applied motivational theories in HCI research~\cite{Tyack2020SDTinGames,Poeller2022SDTLimitations}. which explains how high-quality engagement emerges when activities support people’s basic needs for autonomy, competence, and relatedness. In our context, motivation is not only shaped by game elements such as goals, feedback, and challenge, but also by the specific medium of AR glasses. Therapists framed the glasses as a form of visual augmentation that opens up new ways to structure cognitive–motor rehabilitation tasks—embedding dual-task demands, visual scanning, and spatial cues directly into the patient’s lived environment. In this sense, play is dual: each AR rehabilitation task simultaneously loads the body with specific movement demands and shapes the patient’s psychological and cognitive experience of practice—how motivated, confident, and mentally engaged they feel while moving.
\subsubsection{Engagement}
\begin{table}[h]
\centering
\caption{Subcodes within the Motivation \& Engagement domain.}
\label{tab:ME_subcodes}
\definecolor{dual1}{RGB}{252,208,102}
\par\small\textit{Domain: ME = Motivation \& Engagement. CodeIDs here use the \textbf{ME-} prefix; see Appendix Table~\ref{tab:crosswalk} for the code–domain crosswalk.}
\begin{tabularx}{0.9\linewidth}{l X}
\toprule
\rowcolor{dual1}
\textbf{CodeID} & \textbf{Definition} \\
\midrule
ME-SustainedRepetition & Patients increase practice volume when activities are framed as engaging play. \\
ME-TherapeuticReinforcement & Games provide immediate feedback and encouragement, supporting therapist guidance. \\
ME-FragileEngagement & Motivation is easily disrupted by poor pacing, unclear goals, or excessive stimulation. \\
\bottomrule
\end{tabularx}
\end{table}
PTs consistently highlighted that AR glasses can transform rehabilitation from repetitive exercise into engaging play. Many contrasted traditional therapy, where patients often disengage after a few repetitions, with AR games that sustain attention through novelty, feedback, and challenge. As one PT put it: \emph{Any time you can get a kid doing something that doesn't feel like therapy is better---you have to gamify PT} (PT05). Another noted the motivational layer AR adds to low-tech tasks: \emph{You have lights and colors… it seems very futuristic—more fun than just standing and balancing} (PT09). Others emphasized simple audio feedback for timing and confidence: \emph{Sometimes I heard a ding when I was successful… that was very helpful} (PT03). In this sense, engagement was framed not as superficial entertainment, but as a therapeutic resource-prolonging practice, increasing repetitions, and improving adherence.

Several PTs noted that the game format also reconfigures patient–therapist dynamics. Instead of being the sole source of encouragement, therapists can let the game provide immediate reinforcement, allowing them to focus on observation and adjustment. At the same time, they emphasized that engagement is fragile: overstimulation, unclear goals, or mismatched difficulty could quickly lead to frustration (\emph{There's a little bit of technical frustration there} (PT03);\emph{This game has a bit of a learning curve for everyone} (PT06)). Thus, PTs saw motivation as something to be carefully scaffolded, not automatically guaranteed by the presence of digital play.

Our coding clustered these perspectives into three subcodes (Table~\ref{tab:ME_subcodes}): (1) \emph{Sustained Repetition} (patients perform more repetitions when tasks feel playful), (2) \emph{Therapeutic Reinforcement} (games provide feedback and encouragement alongside the therapist), and (3) \emph{Fragile Engagement} (motivation depends on clarity, pacing, and appropriate challenge).

\subsubsection{Self-Efficacy}

Beyond sustaining motivation, PTs emphasized that AR glasses could play a unique role in helping patients rebuild confidence in their bodies. Many patients arrive in therapy with hesitancy, shaped by pain, prior failures, or fear of re-injury. PTs described how small wins in AR\textemdash{}hitting a target, completing a reach, or balancing through a digital challenge\textemdash{}can generate brief but powerful moments of accomplishment. As one PT noted: \emph{When they succeed in the game, they start to trust their body again} (PT07). Such confidence-building was seen as critical for restoring willingness to engage in progressively harder tasks.

Self-efficacy was not limited to motor success. Several PTs described how playful AR tasks could shift patients’ attention away from symptoms, allowing them to re-experience movement as achievable rather than threatening. \emph{If somebody is afraid of moving, this might be a good distraction to get them moving without really thinking about something being painful or injuring themselves} (PT05). Another explained how AR tasks can offload conscious control: \emph{The task would take their consciousness away from trying to balance. Balance should be unconscious, but after you get hurt, it becomes a conscious thing\textemdash{}this could help transition it back} (PT07). PTs also highlighted uptake among hesitant populations: \emph{Older people who are afraid to move\ldots{} if they can understand the task, they would get into it} (PT08).
\begin{table}[h]
\centering
\caption{Subcodes within the Self-Efficacy domain (confidence restoration and fear reduction).}
\label{tab:SE_subcodes}
\definecolor{dual1}{RGB}{252,208,102}
\par\small\textit{Domain: ME = Motivation \& Engagement. CodeIDs here use the \textbf{SE-} prefix; see Appendix Table~\ref{tab:crosswalk} for the code–domain crosswalk.}
\begin{tabularx}{0.9\linewidth}{l X}
\toprule
\rowcolor{dual1}
\textbf{CodeID} & \textbf{Definition} \\
\midrule
SE-TrustInMovement & Successful task completion helps patients regain belief in their bodily capacity. \\
SE--AttentionalShift & Playful, goal-directed focus reduces symptom vigilance and perceived threat. \\
SE--GradedExposure & Stepwise scaling of range, placement, and repetitions reduces fear of movement. \\
SE--TherapistMediation & PTs pace, guard, and reassure during progression to consolidate confidence. \\
SE--ProgressiveWillingness & Restored confidence enables patients to attempt more challenging movements. \\
\bottomrule
\end{tabularx}
\end{table}

Therapists also framed self-efficacy gains as a mechanism for addressing fear of movement. They emphasized \emph{graded exposure} using AR’s ability to \emph{modulate task difficulty} (e.g., starting with constrained ranges of motion, then gradually expanding target distance, repetition, or spatial placement). Game-based goals\textemdash{}such as catching a digital ball or striking a floating target\textemdash{}recast practice from a pass/fail test into a series of attainable successes, helping patients reinterpret sensations and risk. As one therapist put it: \emph{It’s not just about the muscles\textemdash{}it’s about convincing the patient they can do it again} (PT03). Importantly, PTs described their own ongoing role in pacing, guarding, and reinforcing safe progress, which further stabilized patients’ confidence as tasks scaled up.

\subsubsection{Cognitive Stimulation}

Therapists also emphasized the dual potential of AR games to stimulate both the body and the mind. Many rehabilitation tasks require patients to coordinate motor actions with cognitive processes such as attention, sequencing, or memory. AR glasses, by embedding digital objects into real environments, enable \emph{dual-task training} that combines physical exercise with cognitive engagement. As one PT noted: \emph{It's a nice way to do dual tasking\ldots{} balance while making color associations} (PT01). Another suggested explicit rule overlays: \emph{If there was an ability to call out a number so you had to pick a specific one, that would make it more challenging---it's adding a dual-task component\ldots{} not only do you have to touch it and eliminate it, you now have the additional cognitive task of finding the number first} (PT05).

Participants further highlighted how timing and anticipation can raise cognitive demand without unsafe kinetics. Describing moving targets, a therapist remarked: \emph{Once the gates started to move\ldots{} you have to think about the pace and how they line up\ldots{} that feels like a higher-level processing skill} (PT01). Others pointed to \emph{visual scanning} and spatial attention, especially for neglect: \emph{Awareness of the environment---visual scanning, looking around---could be really good for some people [with inattention/neglect]} (PT05).

We clustered these perspectives into three subcodes (Table~\ref{tab:CS_subcodes}): (1) \emph{Dual-Task Integration} (coordinating motor and cognitive tasks simultaneously), (2) \emph{Sequencing \& Memory} (requiring patients to recall and execute ordered actions), and (3) \emph{Attention \& Distraction Management} (training selective attention by filtering relevant from irrelevant stimuli).

\begin{table}[h]
\centering
\caption{Subcodes within the Cognitive Stimulation (CS) domain. 
CS = Cognitive Stimulation.}
\label{tab:CS_subcodes}
\definecolor{dual1}{RGB}{252,208,102}
\par\small\textit{Domain: COG = Cognitive \& Perceptual. CodeIDs here use the \textbf{CS-} prefix; see Appendix Table~\ref{tab:crosswalk} for the code–domain crosswalk.}
\begin{tabularx}{0.9\linewidth}{l X}
\toprule
\rowcolor{dual1}
\textbf{CodeID} & \textbf{Definition} \\
\midrule
CS-DualTask & Combining movement with concurrent cognitive demands. \\
CS-SequencingMemory & Training memory and sequencing through ordered digital actions. \\
CS-Attention & Strengthening selective attention by managing distractors and task focus. \\
\bottomrule
\end{tabularx}
\end{table}

\section{Discussion}

\subsection{Reframing Personalization and Adaptation: From Auto-Tuning to Therapist Agency}
Prior AR rehabilitation games typically operationalize \emph{personalization} and \emph{adaptation} via (1) \emph{auto-tuning algorithms} (e.g., reinforcement learning that adjusts target placement or step length over time~\cite{Pelosi2024Personalized,alsheikhy2025ml,das2022intelligent,Herrera2025VR}) and (2) \emph{patient-facing} designs that treat patients as the primary end users~\cite{Alamri2010ARRehab,Bank2018Patient,PalaciosNavarro2015Kinect}. While both routes are powerful, uptake can stall when opaque parameter updates and device variability clash with \emph{clinical reasoning} and \emph{patient-centered care}.

Our results suggest a complementary path: \emph{re-center clinicians as decision-makers} and \emph{return adaptation agency to therapists} through lightweight, transparent controls surfaced in the therapist UI. This stance inherits from meta-design and end-user development traditions that position domain experts as end users and co-designers~\cite{FischerGiaccardi2006MetaDesign,Fischer2017MetaDesign,Lieberman2006EndUser,PalomaresPecho2021EndUser}. Concretely, AR games should (1) enable \emph{therapist-bounded spatial authoring} of targets and safe workspaces; (2) expose \emph{editable cognitive layers} (e.g., colors, numeracy, sequence rules) for dual-task demands; and (3) provide \emph{quick presets and on-the-fly adjustments} aligned with session flow, with logs mapped to clinical metrics (e.g., hit rate, tolerated ROM bands, reaction time) and \emph{safety limits} (joint-specific caps, balance risk). Because AR overlays the physical world, these controls can remain \emph{simple and legible} while still enabling rich, therapist-led personalization at the point of care. This sets up our concrete design implications and a worked example.

\subsection{Design Implications}
\subsubsection{PT-Centered Design Guidelines}
Building on the analysis above, we translate our findings into \textbf{eight design implications} for AR glasses rehabilitation—organized around \emph{therapist agency at the point of care}. Each implication is presented through three complementary lenses: (1) the \emph{clinician mental model} that conditions acceptance, (2) the resulting \emph{design guideline} for system behavior and UI, and (3) a \emph{case example} that shows how the principle can play out in situ.

By \emph{clinician mental models} we mean therapists’ working theories about how a tool behaves, what it affords, and how it advances functional goals. In practice, these models crystallize into three adoption checks: \emph{(1)} therapeutic mapping with clear safety bounds (e.g., ROM, symmetry, balance, dual-tasking), \emph{(2)} locus of control during the session (can the therapist bound, sequence, and revise parameters), and \emph{(3)} workflow fit (onboarding, pacing, and documentation). When AR is positioned as an \emph{adjunct} that augments—rather than replaces—clinical judgment, and when controls, presets, and logs are legible to therapists, the system is more readily seen as a partner than a competitor. Table~\ref{tab:design-implications} operationalizes these implications across the three lenses.

\begin{table*}[t]
\centering
\caption{Design Implications for AR Rehabilitation Games.
(Abbrev.: ROM = Range of Motion.)}
\label{tab:design-implications}
\setlength{\tabcolsep}{6pt}
\renewcommand{\arraystretch}{1.15}
\begin{tabularx}{\textwidth}{>{\raggedright\arraybackslash}p{0.18\textwidth} X X X}
\toprule
\textbf{Implication} & \textbf{Clinician Mental Model} & \textbf{Design Guideline} & \textbf{Case Example} \\
\midrule
\textbf{Progressive Postural Flexibility} &
Recovery is staged: posture shifts (sit → stand → balance) must be gradual and safe. &
Provide graded postural modes and tunable ranges; make posture progression explicit and editable. &
Reach game: seated large targets $\rightarrow$ standing midline reaches $\rightarrow$ balance-board weight-shifts with cross-body targets. \\
\addlinespace
\textbf{Therapist-Editable Play Spaces} &
Trust requires visible therapist control; black-box auto-tuning is distrusted. &
Provide drag-to-place target placement, safe-zone definitions, ROM sliders, and symmetry/spatial-region toggles. &Darts game with adjustable target height and restricted region to match patient ROM.\\
\addlinespace
\textbf{Rethinking Difficulty Scaling} &
“Difficulty” is not timing but functional extension (ROM, cross-midline, dual-task). &
Support multi-dimensional scaling: adjust size, number, color; expand spatial envelope; rule complexity; link each dimension to functional outcomes. &
Color-matching reaches: large single-color $\rightarrow$ smaller bi-manual across midline $\rightarrow$ add number-order rules for dual-tasking. \\
\addlinespace
\textbf{Extending PT Toolkits} &
New tools are understood through familiar props (balls, bands, cones, balance boards). &
Build digital twins of PT tools with added metrics (stretch length, dwell time, sway). &
Resistance-band AR exercise with real-time stretch/resistance feedback. \\
\addlinespace
\textbf{Psychological Mediation through Play} &
Fear and self-efficacy shape adherence as much as biomechanics. &
Provide graded exposure, gentle error feedback, and small wins to build confidence. &
Stroke scanning game: start near intact side, expand contralaterally with celebratory feedback \\
\addlinespace
\textbf{Motivational Layering} &
Gamified rewards must be dosed; overstimulation undermines therapeutic clarity. &
Provide adjustable layers of gamification (scores, effects, competition, timing) and a quiet/low-stimulus preset. &
Balance game with toggle for quiet mode vs. competitive scoring. \\
\addlinespace

\textbf{Ball Games as a Design Anchor} &
Ball-based activities are versatile, familiar, and clinically meaningful across specialties. &
Use ML to recognize ball types/goals to create a generalizable rehab grammar. &
Basketball free-throw for ROM; snooker pockets for fine control; golf putting for balance.\\
\addlinespace
\textbf{Onboarding and Cognitive Fit} &
Busy clinicians reject steep learning curves; ease of use is key. &
Minimize onboarding friction; provide clear tutorials that map to existing logics. &
``Three-click start'': posture $\rightarrow$ goal $\rightarrow$ 2-reach calibration; opens with ROM slider + symmetry toggle. \\
\bottomrule
\end{tabularx}
\end{table*}

\subsubsection{Worked Example: PT-in-the-Loop AR Darts}
\begin{figure}[h]
  \centering

  \begin{minipage}[t]{0.23\linewidth}
    \centering
    \includegraphics[width=\linewidth]{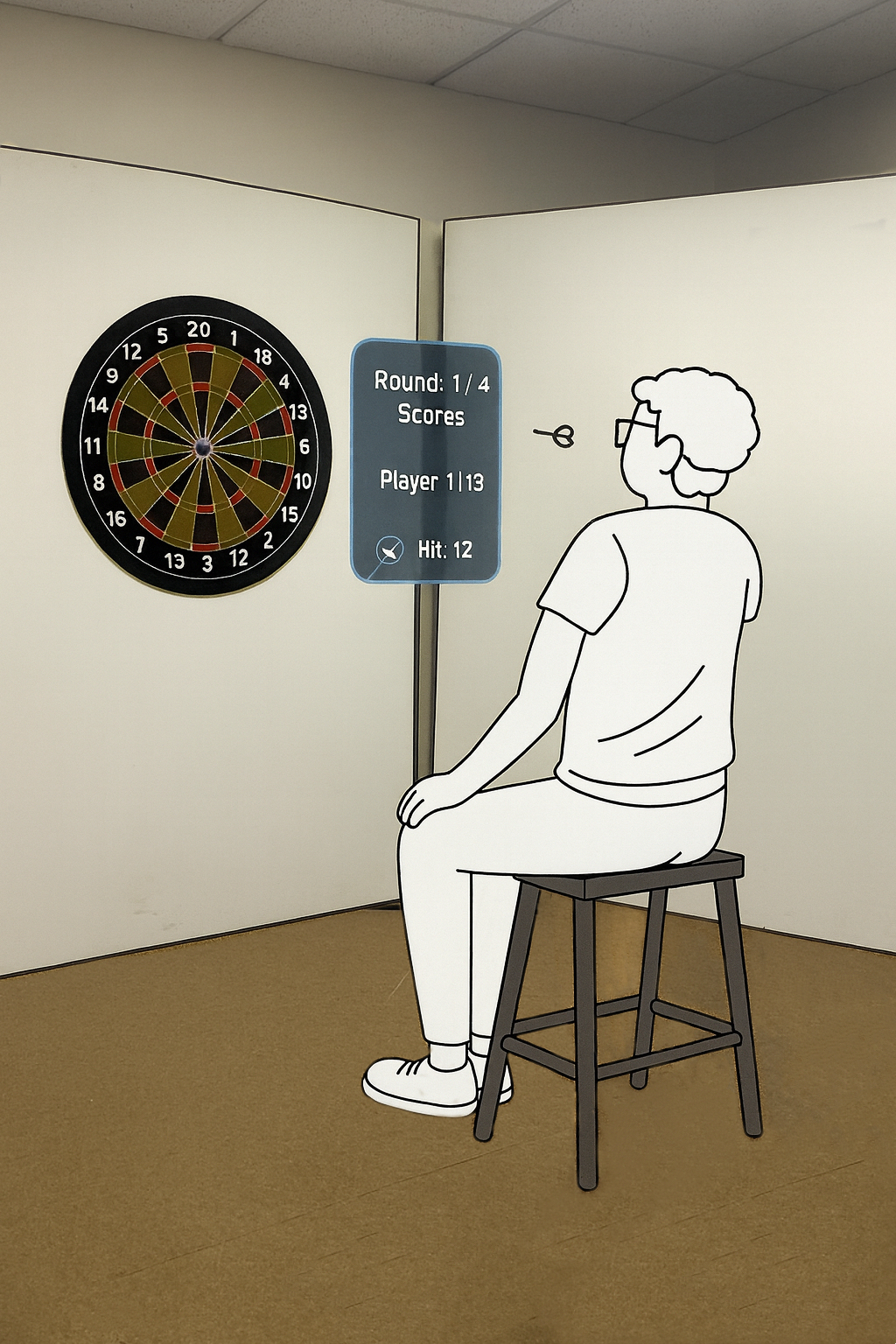}\\[-2pt]
    \small (a)
  \end{minipage}
  \hfill
  \begin{minipage}[t]{0.23\linewidth}
    \centering
    \includegraphics[width=\linewidth]{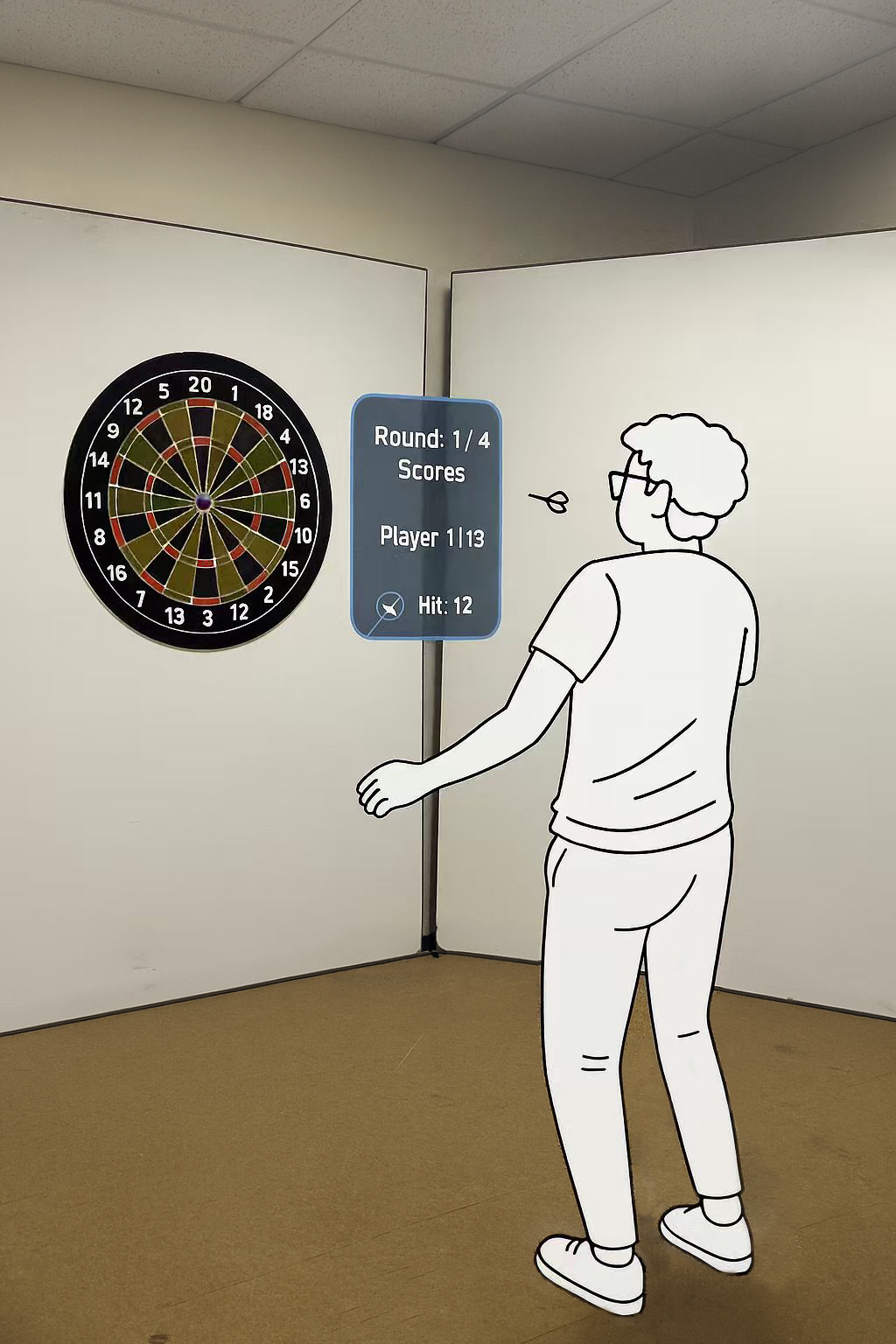}\\[-2pt]
    \small (b)
  \end{minipage}
  \hfill
  \begin{minipage}[t]{0.23\linewidth}
    \centering
    \includegraphics[width=\linewidth]{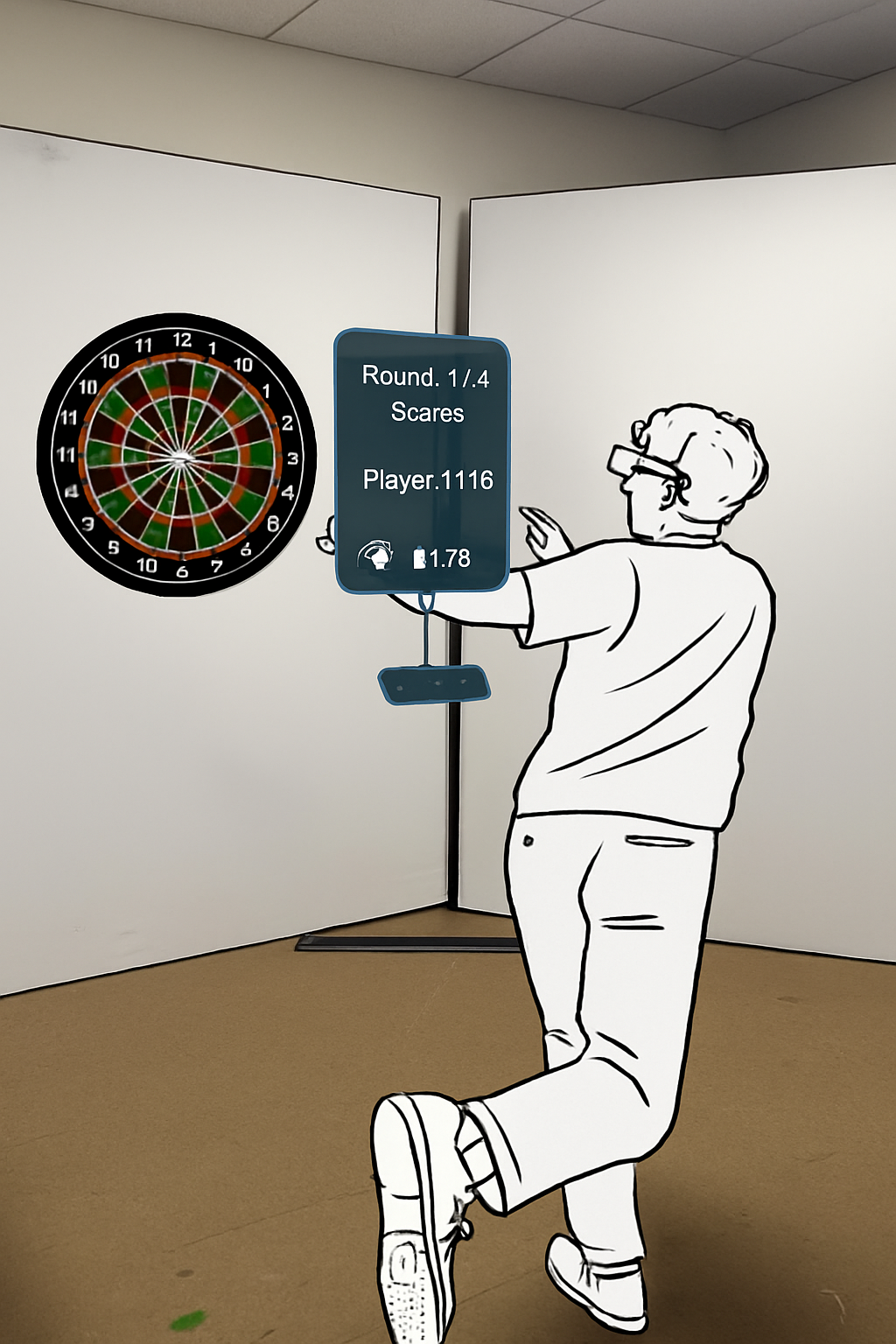}\\[-2pt]
    \small (c)
  \end{minipage}
  \hfill
  \begin{minipage}[t]{0.23\linewidth}
    \centering
    \includegraphics[width=\linewidth]{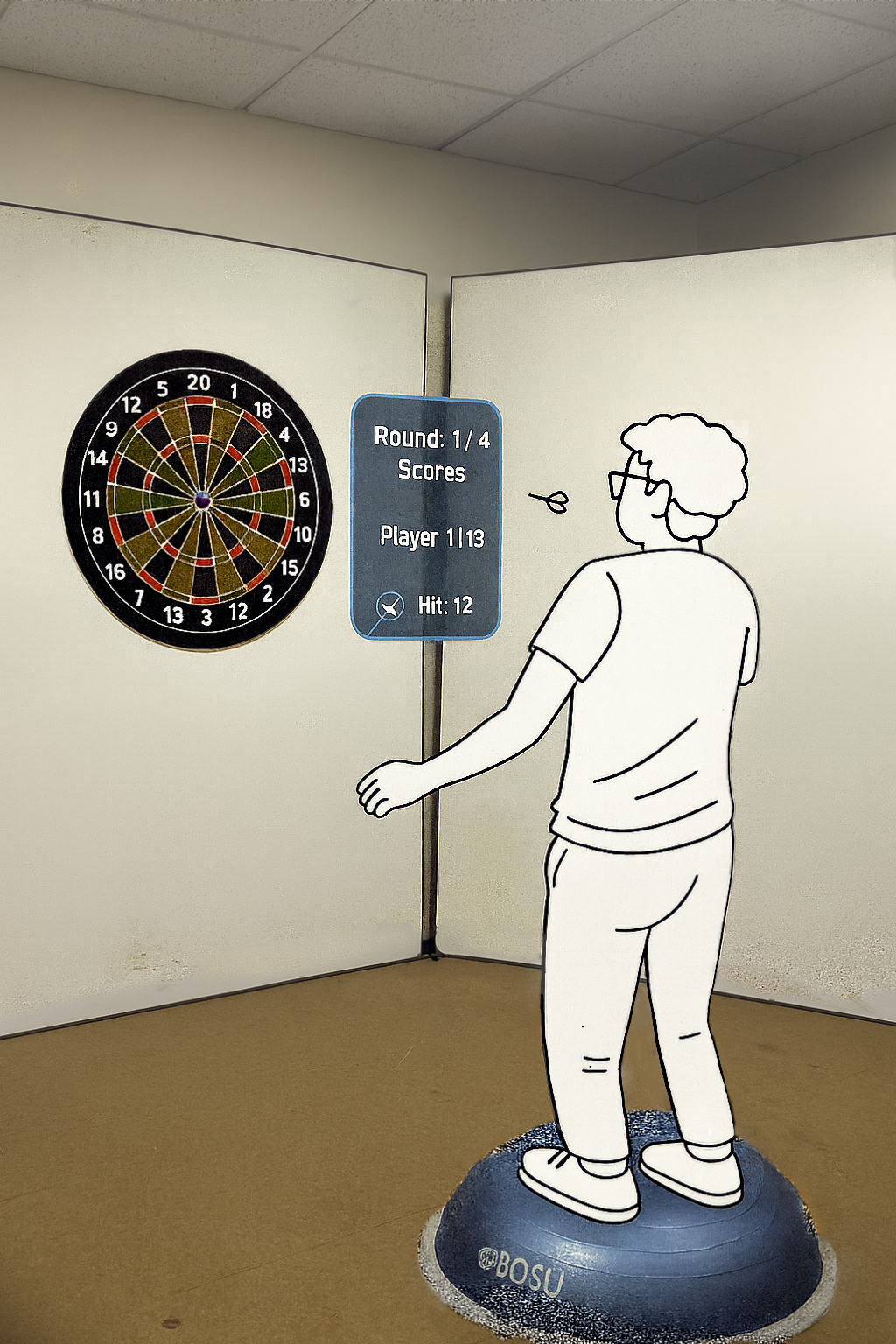}\\[-2pt]
    \small (d)
  \end{minipage}

  \caption{Illustration of the darts case study in AR rehabilitation.
  The same throwing mechanic is graded across postures:
  (a) seated,
  (b) standing with both feet,
  (c) standing on one foot,
  and (d) standing on a balance ball.
  These images focus on postural progression; other variations described in the text (e.g., lunge stance, dual-tasking, peripheral scanning, speed--accuracy trade-offs) are not shown.}
  \label{fig:darts_progression}
  \Description{Four images showing an AR darts game:
  seated,
  standing with both feet,
  standing on one foot,
  and standing on a balance ball.}
\end{figure}

``Darts'' is a familiar therapeutic tool used to target (1) \emph{upper-limb coordination} (shoulder--elbow--wrist sequencing, hand--eye coordination), 
(2) \emph{ROM and strength} (graded shoulder flexion/abduction, forward reach, trunk rotation), 
(3) \emph{posture and balance} (weight shifting, pelvis/trunk control), 
(4) \emph{sensorimotor integration} (vision and proprioception alignment), and (5) \emph{dual-tasking} (color/number matching, serial counting). 
Populations include \emph{neurological} (stroke, TBI, Parkinson’s), \emph{musculoskeletal} (e.g., frozen shoulder; late-phase cuff repair), 
\emph{older adults with balance challenges}, and \emph{pediatrics}. 
In AR, the game can keep the core mechanic identical (aim $\rightarrow$ align $\rightarrow$ release) while exposing a small set of \emph{therapist-editable parameters}---target height, spatial extent/region, sequence rules (colors, numbers), and time limits. 
With only these slots, the same game spans: 
(a) \emph{seated, near-field, large targets} for alignment and gentle release (track hit-rate); 
(b) \emph{standing natural stance} with progressive distance and guided contralateral step for weight shift; 
(c) \emph{lunge/split stance} adding trunk rotation and horizontal abduction control; 
(d) \emph{unstable surfaces} (foam/hemisphere) to emphasize ankle/hip strategies; 
(e) \emph{dual-task layering} (call-and-hit designated numbers or ordered sequences); 
(f) \emph{visual scanning/peripheral engagement} by distributing targets across quadrants; 
and (g) \emph{speed--accuracy trade-offs} under time pressure. 
This vignette illustrates how a \emph{single, consistent mechanic} plus \emph{therapist-controlled progression} can cover heterogeneous needs without proliferating bespoke games.
Figure~\ref{fig:darts_progression} highlights the postural aspects of this case.

\subsection{Limitations}

Our study offers a detailed account of how licensed physical therapists reason about, adapt, and re-author AR rehabilitation games when using lightweight AR glasses in realistic clinical environments. By foregrounding clinical reasoning rather than fixed task performance, the work provides design-oriented hypotheses and a set of practice-grounded levers that can inform future AR rehabilitation systems.

At the same time, our findings are constrained by the absence of patient participants and longitudinal clinical deployment. We therefore cannot make claims about therapeutic effectiveness, adherence over time, or differential impacts across patient populations. Future work must extend these insights into patient-facing trials, long-term use studies, and broader deployment settings to validate and refine the framework proposed here.
\paragraph{PT Attitudes Toward AR}Our aim was to translate PTs’ clinical reasoning into design principles for embodied AR rehabilitation games, rather than to survey general attitudes toward AR. During recruitment, we explicitly introduced lightweight glasses-class AR and framed the study around design principles; consequently, participants were inherently interested in AR game design. Their positive views should not be taken as representative of PTs at large.

\paragraph{Sampling of PT Career Stages}
Our interview sample (\textit{n}=10) skewed toward early-career (1--5 years) and late-career (20+ years) PTs, with only one mid-career (5--20 years) participant. While we sought balance, many mid-career PTs were less available or less interested in discussing AR, plausibly reflecting workload and family responsibilities. Early-career PTs tended to be more comfortable with emerging technologies (often with prior VR experience), whereas experienced PTs—especially those with 30+ years of practice—expressed curiosity about glasses-class AR. These patterns suggest that openness to AR may vary by career stage.

\paragraph{AR Hardware and Cross-Platform Generalizability.}

We conducted the study on Snap Spectacles because they combine a lightweight glasses-class form factor, an on-device display that affords direct embodied interaction, and a comparatively mature developer ecosystem with diverse off-the-shelf AR games—conditions we deemed important for assessing how AR glasses could realistically fit into PT practice. We did not use other candidate platforms: XREAL Air 2/Air 2 Pro require a wired connection to an external device and rely on that device’s input, and Meta Orion was not commercially available during data collection. Spectacles primarily support embodied interaction through computer-vision and IMU-based tracking of body parts, optionally augmented via Bluetooth-connected sensors. Accordingly, our findings are anchored in the capabilities and constraints of this device family. Some implications may transfer under specific conditions—for example, to XREAL when interactions are mediated by IMU-equipped companion devices, or to Meta Orion for body-movement–based interactions—but platforms that incorporate additional sensing modalities (e.g., EMG wristbands and other physiological inputs) may expand the interaction design space and thus limit direct generalizability. Our contribution should be read as a baseline derived from computer-vision/IMU interactions, intended to inform future AR rehabilitation designs as physiological sensing on glasses-class devices matures.

\paragraph{PTs’ Perspective vs. Patients’ Perspective.}
This study centers on PTs’ hands-on playtesting and familiar clinical reasoning to probe how lightweight AR games could be integrated into daily practice. The two-phase design—embodied playtesting followed by semi-structured interviews—was intended to surface movement logics, constraints, and design opportunities as perceived by PTs; accordingly, the framework we derive is PT-informed rather than patient-validated. Because this stage relied on off-the-shelf AR-glasses games developed for general audiences, with immature accessibility features, we did not involve patients—many of whom face mobility, sensory, or cognitive challenges—to avoid avoidable discomfort and misleadingly negative impressions at this early prototype stage. As such, our findings should not be taken to evidence patient usability, acceptability, or therapeutic benefit. Future work will incorporate co-design with therapists and evaluations with diverse patient groups once AR game prototypes are better aligned with accessibility and clinical safety requirements.

\paragraph{United States PT System}
Our study is grounded in the U.S. physical therapy system, where PTs typically hold a Doctor of Physical Therapy (DPT) degree, are licensed at the state level, and in many states practice with direct-access rights and substantial autonomy in clinical decision-making. These professional roles and responsibilities may differ from those of physical therapists in other countries, where training pathways, scopes of practice, and referral requirements can be more constrained or physician-dependent.

\paragraph{Visual Profiles and Optical Compatibility}
None of our clinical experts had prior experience with AR glasses (though some had used VR). Most reported little novelty-related discomfort given the eyeglasses-like form factor, but that same form factor raises practical considerations for vision correction. While common prescriptions were accommodated, one participant with multifocal/\allowbreak progressive lenses experienced modest degradation in clarity when shifting focus. Our findings may therefore underrepresent usability challenges for multifocal prescriptions (and potentially presbyopia or astigmatism). Future work should systematically evaluate optical compatibility (e.g., prescription inserts, adjustable focus/IPD, over-glasses mounting) and recruit broader visual profiles to assess impacts on legibility, depth cues, and task demands.

\subsection{Future Work}
Future work could extend these findings through deployment-focused prototyping with licensed PTs, including iterative in-clinic pilots that refine therapist-facing controls and workflow integration. Subsequent patient-facing studies could examine accessibility, comfort, adherence, and clinical outcomes compared to standard-of-care exercises, as well as evaluate safety limits and outcome sensitivity (e.g., ROM bands, symmetry, dual-task costs). Another direction is the development of lightweight authoring toolkits that enable therapists to configure and share AR game variants across specialties; future systems may also incorporate additional biometrics (e.g., eye tracking, heart rate) to broaden clinical utility.

\section{Conclusion}

This paper reframes lightweight AR glasses as a medium for rehabilitation that must be understood through the lens of clinical reasoning. By examining how licensed physical therapists (PTs) reinterpret and reshape digital play, we show that AR rehabilitation games become meaningful only when aligned with PTs’ situated expertise, treatment goals, and everyday constraints. By foregrounding co-authorship, situatedness, and duality, we illustrate how the clinical refrain “it depends” operates not as an obstacle, but as a generative design principle that guides therapeutic adaptation in practice. Rather than making claims about clinical efficacy, we offer a design-oriented framework to support future systems seeking to align AR rehabilitation games with therapists’ everyday practice. Lightweight AR glasses, we argue, hold promise not because they introduce entirely new forms of play, but because they can extend—and thoughtfully adapt—the tools and logics that therapists already trust.
\begin{acks}
We thank Snap Research for providing Spectacles devices that supported this study, and we are grateful to Jesse McCulloch and Steven Xu for their technical coordination.
We also thank the Northeastern University CHI Crunch community, led by Casper Harteveld, for constructive feedback and academic support throughout the development of this work.
We are grateful to the licensed physical therapists across the United States who generously shared their time and professional expertise.
\end{acks}

\balance
\bibliographystyle{ACM-Reference-Format}
\bibliography{reference} 
\clearpage
\onecolumn
\appendix
\section{Appendix A: Game Selection Materials}
\label{app:game_eval}

\begin{table*}[h]
\centering
\scriptsize
\caption{PT-Oriented Evaluation of 21 Snap AR Games (Rubric a–d)}
\label{tab:pt_eval_21games}
\begin{tabular}{p{2.1cm} p{2.4cm} p{2.8cm} p{2.6cm} p{2.4cm}}
\toprule
\textbf{Game} &
\textbf{Body-Part (a)} &
\textbf{Functional Similarity (b)} &
\textbf{Technical Fidelity (c)} &
\textbf{Clinical Applicability (d)} \\
\midrule

3 Dots--Pinch &
Unilateral UE; fine-motor actions &
Pinch precision; small-range reaching aligned with fine-motor training &
Stable pinch tracking; minor spatial-anchor drift &
Very safe; suitable for seated or standing use; easily graded for different dexterity levels \\

ActionBall &
Bilateral UE involvement &
Reaching and reactive UE control with clear strengthening potential &
Occasional tracking challenge with fast-moving targets &
Standing preferred; seated mode feasible for lower-intensity use \\

ARchER CHAMP &
Bilateral UE; shoulder-dominant &
Graded extension; controlled reaching motion &
Extended reach increases the chance of drift, limiting precise control &
Standing; mild supervision recommended; may not suit patients with limited shoulder ROM or balance \\

Ball Games &
Bilateral LE; repetitive kicking &
Hip/knee extension; sport-like kicking patterns supporting lower-limb and sports-injury rehab &
Easy to use when paired with a real ball &
Requires open space; well-suited for lower-limb and sports-injury rehab \\

Basketball Training &
Whole body; UE+LE involvement &
Agility; coordinated sport-based drills with mostly non-interactive AR content &
Fast motion may exceed tracking capability, and many elements play more like a video than an interactive AR task &
Standing; moderate exertion level; limited AR interactivity (similar to following an exercise video) \\

Sign of Dorm &
UE + trunk rotation &
Bimanual coordination and reaction timing &
Fast gesture switching stresses tracking  &
Standing; moderate balance needed;  \\

Darts &
Unilateral UE; wrist/elbow control &
Precision aiming and controlled extension; good for graded ROM and motor control &
Stable for slow throws; less accurate when the hand passes near or behind the glasses &
Safe for seated or standing users; easy to adapt to varied ability levels \\

LEGO Bricktacular &
Unilateral UE; fine-motor finger use &
Dexterity; grasp--release sequencing closely mirroring fine-motor rehab tasks &
Stable tracking due to small movements &
Very safe; ideal for seated users and fine-motor training \\

Squishy Run &
Whole-body walking with UE reach &
Dual-task gait; dynamic balance with functional walking demands &
Turning and walking may cause drift but task goals remain understandable &
Clear walkway needed; supervision advised; strong candidate for dual-task gait training \\

Beat Boxer &
UE + weight-shift engagement &
Rhythmic reaching; upper-limb strengthening with strong engagement potential &
Fast strikes challenge tracking robustness but core targets remain clear &
Standing; supervision recommended, especially for higher-intensity play \\

Peridot Beyond &
UE gestures; minimal trunk use &
Gentle reach-to-touch motions with limited functional load &
Stable for slow gestural input &
Safe seated or standing; lower intensity and narrower therapeutic focus \\

Path Pioneer &
Walking-focused; LE dominant &
Endurance gait and navigation in less structured environments &
Outdoor GPS/AR anchor variability &
Outdoor safety considerations important; clinic use constrained by context and weather \\

Pop Game &
Walking with reaching tasks &
Dual-task gait; reactive UE reaching with clear, repeatable targets &
Turning motions introduce drift risk but the core task remains interpretable &
Open space needed; optional supervision; good candidate for combined gait and UE training \\

Ice Fishing (Mobile) &
Unilateral UE; wrist deviation &
Wrist mobility; fine motor control in a narrow movement range &
Phone-based interaction &
Safe seated or standing; limited whole-body or balance involvement \\

Zombie Chase &
Whole-body rapid stepping &
Agility with multidirectional stepping &
High-motion conditions reduce tracking stability and feedback reliability &
High exertion; supervision recommended; intensity may exceed many rehab scenarios \\

Whack-a-Mole &
Bilateral UE; full-arm swings &
Large-arc reaching; UE power tasks with meaningful ROM demands &
Vertical swings tracked reliably &
Safe seated or standing; low risk and simple to explain \\

Golf &
UE + trunk rotation &
Controlled swing; graded force modulation but sport-specific &
Phone-based interaction &
Safe for most users; seated optional; technique demands may limit some patient groups \\

S-CAB &
UE pinch + trunk lean &
Trunk stability via simultaneous trunk lean and pinching, with high coordination demands &
Large reaches can induce drift and hinder precise grading &
Standing; seated variation possible; complex body–finger coordination \\

Rider Rush SE &
UE steering + trunk lean &
Coordination; similar in complexity to S-CAB &
Similar tracking issues as S-CAB &
Similar coordination difficulty  \\

Tiny Hoops &
UE pinch + release &
Grasp--release; target-oriented reaching in a casual context &
Stable pinch tracking &
Repetitive fast pinching may cause fatigue  \\

Pac-Man Fit &
Walking + UE punching/dodging &
Dynamic gait; directional balance with frequent rapid movements &
High movement increases anchor drift and can disrupt consistent feedback &
Requires space; supervision for safety; higher intensity than many standard rehab use-cases \\

\bottomrule
\end{tabular}

\vspace{4pt}
\textbf{Abbreviations:} UE = Upper Extremity; LE = Lower Extremity; ROM = Range of Motion; Drift = AR anchor misalignment.
\end{table*}

\clearpage
\twocolumn
\section{Appendix B: Game Selection Study Protocol}
\label{app:pilot-protocol}

\subsection{Section 1: Pre-Study Survey (Therapist Background Information)}
Participants completed a brief background questionnaire prior to the AR game playtesting session.

\begin{enumerate}
    \item \textbf{Age} \\
    \(\square\) Under 25 \quad
    \(\square\) 25--34 \quad
    \(\square\) 35--44 \quad
    \(\square\) 45--54 \quad
    \(\square\) 55--64 \quad
    \(\square\) 65+
    
    \item \textbf{Gender} \\
    \(\square\) Female \quad
    \(\square\) Male \quad
    \(\square\) Non-binary / third gender \quad
    \(\square\) Prefer not to say
    
    \item \textbf{Clinical specialty / subspecialty} \\
    (e.g., neurorehabilitation, orthopedics, geriatrics, pediatrics, sports medicine) \\
    \underline{\hspace{0.9\linewidth}}
    
    \item \textbf{Years of practice as a PT} \\
    \(\square\) Less than 1 year \quad
    \(\square\) 1--3 years \quad
    \(\square\) 4--6 years \quad
    \(\square\) 7--10 years \quad
    \(\square\) 11--15 years \quad
    \(\square\) More than 15 years
\end{enumerate}

\subsection{Section 2: Introduction and Warm-Up}
Before starting the main tasks, the researcher:
\begin{itemize}
    \item introduced the overall study structure, including that participants would playtest nine AR games across four PT-relevant categories, be encouraged to think aloud while playing, answer brief questions after each game, and respond to a few cross-game summary questions at the end; and
    \item guided the participant through a short pre-game warm-up using the ``Tutorial'' lens in the Spectacles library to familiarize them with AR interaction and basic device operation.
\end{itemize}

\subsection{Section 3: Playtesting Session}
Participants were asked to play games from each PT-relevant category. The order of the four categories was randomized, and the order of games within each category was randomized as well. Participants could decide how long to play each game and could stop at any time.

\begin{table}[t]
\scriptsize
\centering
\caption{PT-informed grouping and within-group ordering of the selected Snap AR games.}
\label{tab:pt_game_groups}

\setlength{\tabcolsep}{4pt}
\renewcommand{\arraystretch}{0.95}

\begin{tabularx}{\columnwidth}{
  >{\raggedright\arraybackslash}p{2.6cm}
  >{\raggedright\arraybackslash}X
}
\toprule
\textbf{PT-Relevant Category} & \textbf{Games (preferred $\rightarrow$ backup)} \\
\midrule
Upper Extremity (UE) & ActionBall; Beat Boxer; Whack-a-Mole \\
Whole Body & Squishy Run; Pop Game \\
Neuro & 3 Dots--Pinch; LEGO Bricktacular \\
Real-Tool Integration & Ball Games; Darts \\
\bottomrule
\end{tabularx}
\end{table}

After each game, participants completed a short rating sheet covering the following dimensions:

\begin{itemize}
    \item Target body area (e.g., arm, leg, balance)
    \item Repetitive movement suitability (1--5)
    \item Interactive feedback quality (1--5)
    \item Patient engagement potential (1--5)
    \item Safety / fatigue risk (1--5)
    \item Recommended user type (e.g., stroke, older adult, child)
    \item Comments / notes (free text)
\end{itemize}

\subsection{Section 3A: Follow-Up Interview Questions (After Each Game Category)}
After completing each game category, participants were asked semi-structured follow-up questions.

\paragraph{Movement relevance}
\begin{itemize}
    \item What movements in the game seem therapeutically useful?
    \item Are these similar to movements you typically use in therapy?
    \item Are there any movements that may be risky or inappropriate for some patients?
\end{itemize}

\paragraph{Engagement and cognition}
\begin{itemize}
    \item What cognitive function(s) does this game engage or relate to?
\end{itemize}

\paragraph{Clinical suitability}
\begin{itemize}
    \item What kinds of patients could benefit from this game?
    \item At which stage of rehabilitation might you use it?
    \item What changes would improve its clinical applicability?
\end{itemize}

\subsection{Section 3B: Cross-Game Comparison (After All Games)}
\paragraph{Cross-game comparison and general feedback}
\begin{itemize}
    \item Which game has the strongest rehabilitation potential? Why?
    \item Which game seemed least applicable? Why?
    \item Were any key therapy elements missing across these games?
    \item What core features would you prioritize in a new AR rehabilitation game?
\end{itemize}

\paragraph{Closing: Future involvement}
\begin{itemize}
    \item Would you be open to future playtesting or providing additional design input?
    \item Do you have colleagues who might be interested in joining future studies?
\end{itemize}

\clearpage
\section{Appendix C: Main Study Protocol}
\label{app:main-protocol}

\subsection{Section 1: Pre-Study Survey (Main Study Participant Background)}
Participants completed a brief background questionnaire prior to the main AR game evaluation session.

\begin{enumerate}
    \item \textbf{Years of experience in physical therapy} \\
    \underline{\hspace{0.3\linewidth}}
    
    \item \textbf{Primary area(s) of clinical expertise} \\
    (e.g., orthopedics, geriatrics, neurology, sports medicine, pediatrics, cardiopulmonary, women’s health, vestibular) \\
    \underline{\hspace{0.9\linewidth}}
    
    \item \textbf{Experience with digital or interactive rehabilitation tools} \\
    Have you used or prescribed any of the following in your practice? (Check all that apply.) 
    \begin{itemize}
        \item \(\square\) Virtual reality rehab systems (e.g., Oculus-based therapy)
        \item \(\square\) Motion-tracking rehab (e.g., Kinect, Wii Fit)
        \item \(\square\) Mobile apps for home exercise or feedback
        \item \(\square\) Wearable sensors or smart garments
        \item \(\square\) Other: \underline{\hspace{0.45\linewidth}}
        \item \(\square\) None
    \end{itemize}
    
    \item \textbf{Familiarity with augmented reality (AR) technologies} \\
    How would you describe your familiarity with AR (e.g., HoloLens, Spectacles, Magic Leap)?
    \begin{itemize}
        \item \(\square\) Never used or seen AR in person
        \item \(\square\) Somewhat familiar (e.g., seen demos or videos)
        \item \(\square\) Used AR once or twice
        \item \(\square\) Regular user or have integrated AR into practice/research
    \end{itemize}
    
    \item \textbf{Familiarity with virtual reality (VR) technologies (optional)} \\
    How would you describe your familiarity with VR (e.g., Oculus/Meta Quest, HTC Vive)?
    \begin{itemize}
        \item \(\square\) Never used or seen VR in person
        \item \(\square\) Somewhat familiar (e.g., seen demos or videos)
        \item \(\square\) Used VR once or twice
        \item \(\square\) Regular user or have integrated VR into practice/research
    \end{itemize}
\end{enumerate}

\subsection{Section 2: Study Overview and Instructions}
Before starting the main AR game playtesting, the researcher:
\begin{itemize}
    \item introduced the overall study structure, including that participants would playtest four pre-selected AR mini-games while wearing the AR glasses, be encouraged to think aloud during gameplay, answer brief reflection questions after each game, and respond to several cross-game summary questions at the end; and
    \item demonstrated the basic operation of the AR glasses (e.g., how to wear the device, start and exit games, and maintain safety during movement) without using a dedicated tutorial lens.
\end{itemize}

\subsection{Section 3: AR Game Playtesting Session}
Physical therapists (PTs) played four pre-selected AR mini-games while wearing the AR glasses.

During gameplay:
\begin{itemize}
    \item participants were encouraged to think aloud, commenting on movement comfort, physical engagement, and safety.
\end{itemize}

After each game:
\begin{itemize}
    \item participants verbally reflected on which patient types might benefit from or struggle with the game; and
    \item discussed potential modifications to improve the game's clinical relevance and usability.
\end{itemize}

\subsection{Section 3A: Semi-Structured Interview}
After each game, the researcher used a semi-structured interview guide to prompt reflection.A seat was provided, and participants could choose whether to sit or stand.
\paragraph{Movement relevance}
\begin{itemize}
    \item What movements in the game seem therapeutically useful?
    \item Are these similar to movements you typically use in therapy?
    \item Are there any movements that may be risky or inappropriate for some patients?
\end{itemize}
\paragraph{Engagement and cognition}
\begin{itemize}
    \item Do you think this game is related to any cognitive function(s)? If so, which ones?
\end{itemize}
\paragraph{Clinical suitability}
\begin{itemize}
    \item What kinds of patients could benefit from this game?
    \item At which stage of rehabilitation might you use it?
    \item What changes would improve its clinical applicability?
\end{itemize}

\subsection{Section 3B: Cross-Game Comparison and General Feedback}
After completing all four games, participants were asked to compare across games and provide overall feedback.

\paragraph{Cross-game comparison and general feedback}
\begin{itemize}
    \item Which game has the strongest rehabilitation potential? Why?
    \item Which game seemed least applicable? Why?
    \item Were any key therapy elements missing across these games?
    \item What core features would you prioritize in a new AR rehabilitation game?
\end{itemize}

\clearpage
\onecolumn

\section{Appendix D: Crosswalk from Subcodes to the Six Analytic Domains}
\begin{table*}[h]
\centering
\caption{Crosswalk from subcodes to the six analytic domains. 
Domains: CA = Clinical Alignment; COG = Cognitive \& Perceptual; RS = Risk \& Safety; TF = Technology \& Feedback; ME = Motivation \& Engagement; DP = Design Principles.}
\label{tab:crosswalk}
\small
\begin{tabularx}{\textwidth}{l c X}
\toprule
\textbf{CodeID} & \textbf{Primary Domain} & \textbf{Rationale (one-line)} \\
\midrule
CA-Func & CA & Maps digital tasks to functional outcomes (ROM, balance). \\
CA-ADL & CA & Links mechanics to activities of daily living. \\
CA-Pop & CA & Targets specific patient populations (pediatrics, stroke). \\
CA-BodyPart & CA & Explicit body-region alignment (shoulder, trunk, lower limb). \\
CAA-RangeCalibration & CA & Range set by condition-specific limits (e.g., frozen shoulder). \\
CAA-JointSafety & CA & Posture/loading safety for affected joints. \\
CAA-AsymmetryCompensation & CA & Biasing targets to weaker/neglected side. \\
\midrule
PMD-PreCalibration & DP & Pre-session setup reflecting clinical reasoning (design-time choice). \\
PMD-InPlayAdaptation & DP & Therapist-driven, on-the-fly parameter edits (design control). \\
PMD-Posture & DP & Structured posture ladder (seated $\rightarrow$ standing $\rightarrow$ single-leg). \\
PMD-Complexity & DP & Sequencing/dual-task/cross-body as planned progression axes. \\
PMD-Load & DP & Progressing stability/load via surfaces/props per design plan. \\
\midrule
CDC-Target Placement & DP & Therapist places/weights spatial regions (authoring control). \\
CDC-Stimulus Scaling & DP & Size/frequency/salience as editable content knobs. \\
CDC-Therapist Control & DP & Explicit priority-setting by therapist over game defaults. \\
\midrule
AD-TargetParams & TF & System-level tuning of target properties for difficulty. \\
AD-SpatialEnvelope & TF & System expands/shifts FOV within configured bounds. \\
AD-TemporalPattern & TF & Temporal scheduling beyond speed handled by engine. \\
AD-CognitiveLayering & TF & Rule overlays (numbers, sequencing) implemented in logic. \\
AD-SafetyAssist & TF & Hitbox/magnetism/auto-ease are feedback/tuning features \emph{(RS cross-domain)}. \\
\midrule
SSSA-EarlyStage & RS & Guarded, low-intensity use in ICU/acute settings. \\
SSSA-LateStage & RS & Higher-intensity/dual-task use in outpatient/sports. \\
SSSA-Environment & RS & Space/hygiene/supervision constraints and setup. \\
\midrule
DPT-VirtualizedProps & TF & Virtual props implemented by system (balls, cones, bands). \\
DPT-AllInOne & TF & Reduces physical equipment burden via software substitutes. \\
DPT-AugmentedFeedback & TF & Adds visual/audio/adaptive feedback unavailable physically. \\
\midrule
ME-SustainedRepetition & ME & Play sustains volume of practice. \\
ME-TherapeuticReinforcement & ME & Feedback/reinforcement complements therapist. \\
ME-FragileEngagement & ME & Engagement depends on clarity/pacing/challenge. \\
\midrule
SE-TrustInMovement & ME & Small wins rebuild confidence in capability. \\
SE--AttentionalShift & ME & Focus on goals reduces symptom vigilance. \\
SE--GradedExposure & ME & Stepwise scaling reduces fear of movement. \\
SE--TherapistMediation & ME & Pacing/guarding/reassurance consolidate confidence. \\
SE--ProgressiveWillingness & ME & Renewed confidence enables harder tasks. \\
\midrule
CS-DualTask & COG & Concurrency of motor and cognitive demands. \\
CS-SequencingMemory & COG & Ordered actions and recall. \\
CS-Attention & COG & Selective attention and distractor management. \\
\bottomrule
\end{tabularx}
\end{table*}
\twocolumn

\begin{table*}[t]
\small
\caption{Traceability from analytic domains to interpretive lenses.}
\label{tab:traceability}
\begin{tabularx}{\textwidth}{p{0.18\textwidth} Y Y Y}
\toprule
\textbf{Domain} & \textbf{Play is Co-Authored} & \textbf{Play is Situated} & \textbf{Play is Dual} \\
\midrule
\textbf{CA Clinical Alignment} 
& \makecell[l]{Movement--body logic;\\ therapist-authored\\ functional goal mapping} 
& \makecell[l]{Condition-/stage-/setting-\\ specific bounds on play} 
& \textcolor{gray}{(not primary)} \\

\textbf{COG Cognitive \& Perceptual} 
& \textcolor{gray}{(not primary)} 
& \makecell[l]{Perceptual/cognitive load\\ as contextual constraint} 
& \makecell[l]{Dual-task and cognitive\\ engagement during movement} \\

\textbf{RS Risk \& Safety} 
& \textcolor{gray}{(not primary)} 
& \makecell[l]{Safety thresholds;\\ setting- and stage-\\ sensitive constraints} 
& \makecell[l]{Psychological safety /\\ avoidance concerns} \\

\textbf{TF Technology \& Feedback} 
& \makecell[l]{Therapist-facing controls;\\ non-speed difficulty tuning} 
& \makecell[l]{Digitizing PT toolkits;\\ environment-aware feedback} 
& \textcolor{gray}{(not primary)} \\

\textbf{ME Motivation \& Engagement} 
& \textcolor{gray}{(not primary)} 
& \textcolor{gray}{(not primary)} 
& \makecell[l]{Enjoyment, adherence,\\ self-efficacy, fear reduction} \\

\textbf{DP Design Principles} 
& \makecell[l]{Progressive movement design;\\ customizable content} 
& \makecell[l]{Parameter bounds across\\ contexts} 
& \makecell[l]{Goal/reward meaning} \\
\bottomrule
\end{tabularx}
\end{table*}

\end{document}